\documentclass[a4paper,fleqn,usenatbib,useAMS]{mnras}

\usepackage{newtxtext,newtxmath}

\usepackage[T1]{fontenc}
\usepackage{ae,aecompl}

\usepackage{graphicx}	
\usepackage{amsmath}	
\usepackage{amssymb}	
\usepackage{natbib}
\usepackage{multirow}
\usepackage{pdflscape}

\usepackage{color}
\usepackage[normalem]{ulem}

\newcommand{\fap}{\mathrm{FAP}}	

\title[Secondary modes of classical Cepheids]{Investigating light curve modulation via kernel smoothing. \\
II. New additional modes in single-mode OGLE classical Cepheids}

\author[M.~S\"uveges and R.I.~Anderson]{
Maria S\"uveges$^{1}$\thanks{E-mail: sueveges@mpia.de} and
Richard I. Anderson$^{2}$\thanks{ESO fellow, E-mail: randerso@eso.org}
\\\\
$^{1}$ Max Planck Institute for Astronomy, K\"onigstuhl 17, 69117 Heidelberg, Germany\\
$^{2}$ European Southern Observatory, Karl-Schwarzschild-Str. 2, D-85748 Garching b. M\"unchen, Germany}

\date{Accepted XXX. Received YYY; in original form ZZZ}

\pubyear{2017}

\begin{document}

\defcitealias{suvegesanderson17}{Paper I}

\label{firstpage}
\pagerange{\pageref{firstpage}--\pageref{lastpage}}
\maketitle

\begin{abstract}
Detailed knowledge of the variability of classical Cepheids, in particular their modulations and mode composition, provides crucial insight into stellar structure and pulsation. However, tiny modulations of the dominant radial-mode pulsation were recently found to be very frequent, possibly ubiquitous in Cepheids, which makes secondary modes difficult to detect and analyse, since these modulations  can easily mask the potentially weak secondary modes. The aim of this study is to re-investigate the secondary mode content in the sample of OGLE-III and -IV single-mode classical Cepheids using kernel regression with adaptive kernel width for pre-whitening, instead of using a constant-parameter model. This leads to a more precise removal of the modulated dominant pulsation, and enables a more complete survey of secondary modes with frequencies outside a narrow range around the primary. Our analysis reveals that significant secondary modes occur more frequently among first overtone Cepheids than previously thought. The mode composition appears significantly different in the Large and Small Magellanic Clouds, suggesting  a possible dependence on chemical composition. In addition to the formerly identified non-radial mode at $P_2 \approx 0.6\ldots 0.65 P_1$ (0.62-mode), and a cluster of modes with near-primary frequency, we find two more candidate non-radial modes. One is a numerous group of secondary modes with $P_2 \approx 1.25 P_1$, which may represent the fundamental of the 0.62-mode, supposed to be the first harmonic of an $l \in \{7,8, 9\}$ non-radial mode. The other new  mode is at $P_2 \approx 1.46 P_1$, possibly analogous to a similar, rare mode recently discovered among first overtone RR Lyrae stars. 
\end{abstract}




\section{Introduction}
\label{sec:intro}

Classical Cepheids (henceforth: Cepheids) have played and continue to play a key role for testing stellar models. For instance, Cepheid properties are highly sensitive to several key ingredients of stellar evolution models, such as convective core overshooting, mass loss, and rotational mixing \citep{bonoetal2000,neilson+langer08,andersonetal14,andersonetal16b}. At the same time, the large amplitude Cepheid light and radial velocity variations continue to provide challenges for hydrodynamical models \citep{nataleetal2008,marconietal13,nardettoetal17,vasilyevetal17}. Compared to the rich multi-mode phenomena seen in $\delta$ Scuti stars or the solar-like oscillations detected in red giant stars, the variability of Cepheids has thus far seemed very simple. However, recent studies involving highly precise photometric data and/or particularly long temporal baselines have revealed previously hidden complexity of Cepheid pulsations. Given the immense success of asteroseismology and the importance of Cepheids for stellar physics, the possibility that Cepheids may have additional, possibly non-radial, modes that can be exploited by asteroseismic methods is an exciting prospect that deserves further attention.
 
The extensive sample of Cepheids observed by the Optical Gravitational Lensing Experiment \citep[OGLE]{udalskietal97, soszynskietal08, udalskietal15} provides high-quality photometric data with sufficient temporal baselines for identifying possible secondary and further modes in classical Cepheids and to test for the occurrence rate of such additional variability. 
Analyses by e.g. \citet{moskalikkolaczkowski09, soszynskietal10, soszynskietal15a} of OGLE data have revealed the simultaneous presence of two or three radial modes in some Cepheids and of non-radial modes in others.

However, several studies have shown that, in addition to the secular period changes known since long \citep{szabados83, berdnikovignatova00, pietrukowicz01, pietrukowicz02,  poleski08, turneretal06}, fluctuations of pulsation periods and amplitudes are present in some, if not all, Cepheids. An early example of this was V473\,Lyr, which exhibits periodic variations of its light and radial velocity amplitude, similar to RR Lyrae stars exhibiting the Blazhko effect \citep{burkietal82,molnaretal13}. Space-based photometry obtained with the {\it Kepler}, {\it MOST}, and {\it BRITE} satellites has revealed a host of variations in the fundamental-mode Cepheid V1154\,Cyg \citep{derekasetal12,kanevetal15, derekasetal17}, found long-period modulation in T\,Vul \citep{smolecetal16} and  suggested a greater degree of irregularity exhibited by first-overtone (FO) Cepheids \citep{evansetal15}. \citet{smolec17} detected strictly periodic modulations in the pulsation periods of 51 Magellanic Cloud Cepheids using OGLE data. 
Most recently, relaxing the condition of periodicity, we have presented a new methodology for uncovering modulated variability and applied it to OGLE photometry of 53 LMC Cepheids \citep[][hereafter Paper~I]{suvegesanderson17}. We have revealed variations of period and amplitude, composed of irregular, multi-scale or stochastic periodic and slow trend-like components. The wide range of modulations  suggests these phenomena to be ubiquitous and their detectability to be limited by the time sampling and the photometric precision available.

Previous investigations of additional pulsation modes have applied constant-parameter Fourier analysis to pre-whiten time series and search for new significant periodicities in the residuals \citep{moskalikkolaczkowski09, soszynskietal10, soszynskietal15a, smolecsniegowska16}. As discussed in \citetalias{suvegesanderson17}, time-dependent phenomena in the pulsation parameters can decrease the performance of the canonical constant-parameter pre-whitening methods, since fitting a fixed light curve pattern using a constant period leaves unexplained variations in the light curve residuals, increasing the detection limit of weak secondary modes.

The aim of the present work is to carry out a more sensitive search for weak (possibly non-radial) secondary modes in single-mode classical Cepheids using the latest available OGLE photometry of both Magellanic Clouds. Instead of a constant-parameter pre-whitening, we apply the local kernel regression \citep{fangijbels} method developed in \citetalias{suvegesanderson17} for pre-whitening, which traces the time-varying light curve parameters across a series of short sliding windows using a weighting scheme that lends stronger influence to observations close to the window centre. The result is a more precise estimate of the instantaneous brightness of the star at any given window centre than can be obtained by a constant model. This increases the chances for finding weak periodic signals in the residuals.



We summarize the used data in Section \ref{sec:data} and the employed statistical methodology, namely, the local kernel regression and the significance assessment in Section \ref{sec:methods}. In Section \ref{sec:meth_simulations}, we discuss the efficiency of this procedure in detecting small-amplitude signals in the presence of a strong modulated primary signal, and determine the range of frequencies which can be recovered, using simulations of weak signals added to noisy, modulated dominant modes. After a general accounting of the  significant secondary modes as a function of Cloud and mode, Section \ref{sec:results} discusses the modes one by one. In Section \ref{sec:char}, we characterise the distribution of their amplitudes, the primary periods at which they occur,  and their position in the colour-magnitude diagram, comparing them by mode and by Cloud. Section \ref{sec:concl} summarizes our findings.

\section{Data}
\label{sec:data}

Our study uses the OGLE-II, OGLE-III and OGLE-IV I-band photometric time series data of classical Cepheids. Observations were taken in 1997--2000 \citep[OGLE-II,][]{udalskietal97}, in 2001--2009 \citep[OGLE-III,][]{soszynskietal08} and in 2010--2015 \citep[OGLE-IV, ongoing][]{udalskietal15}, with two long gaps (up to about 300 and 700 days, resp.) and instrumentation changes between the three phases of the survey. Observations were conducted using the 1.3-meter Warsaw telescope at Las Campa\~nas Observatory in Chile. The data were reduced using the Difference Image Analysis technique \citep{alardlupton98, wozniak00, udalskietal08}.The second and third survey phases were cross-calibrated  using about 600,000 stars from the overlapping regions \citep{udalskietal08}. OGLE-IV photometry was calibrated using thousands of OGLE-III photometric standard stars in fields overlapping with OGLE-III maps \citep{udalskietal15}. The time sampling is seasonal, with yearly gaps of about 100 days. Typical error bars on the I-band photometry are below 0.02 mag. The OGLE-III and -IV catalogs of variable stars provide separate periods for each Cepheid. Where available, we use the non-weighted average of these as the initial value for optimisation procedures and to compute period ratios. Otherwise, we used the single available value.

All data were obtained from the OGLE website\footnote{\texttt{http://ogle.astrouw.edu.pl}}. 

\section{Methods}
\label{sec:methods}

\subsection{Local kernel regression}
\label{subsec:kernel}

In \citetalias{suvegesanderson17}, we found that tiny but perceptible modulations of period and harmonic amplitudes may be ubiquitous in single-mode Cepheids. Such an instability of the primary pulsation hampers the standard constant-parameter pre-whitening technique, which was used for the frequency analysis of OGLE data \citep[see, e.g.,][]{soszynskietal15b, smolecsniegowska16, smolec17}. The unaccounted-for modulations contribute to the noise, and make harder to detect small-amplitude second modes, compromising thus their parameter estimates and population statistics. 

To improve on this issue, we applied the local kernel method described in \citetalias{suvegesanderson17}  to obtain better fits of the primary pulsation and more precise fitted magnitude values at the observational epochs. We refer to \citetalias{suvegesanderson17} for a detailed description of the method adapted to astronomical time series, and to \citet{fangijbels} for an introduction to local regression in statistics.
Briefly summarised, the procedure consists of fitting a harmonic series plus a low-order polynomial trend
in a sequence of time windows, which are centred on a grid of time points (the order of the harmonic series is taken from stable-parameter models individually for each star as described below in Section \ref{subsec:input}). The harmonic and polynomial parameters as well as the period are treated as unknown and estimated by weighted nonlinear least-squares regression in each window. Thus, the series of estimated periods and harmonic parameters at the gridpoints follow the temporal variations of these parameters. To obtain smooth curves, and to decrease the bias of the estimates, the product of a Gaussian kernel and the usual inverse error weights is used for weighting. The Gaussian component downweights the effect of the observations far from the centre, and thus decreases the bias and reduces artificial jumps in the estimates when influential observations enter or leave the window. The inverse error weights, as usual, weaken the effect of comparatively noisy observations on the estimates. 

The method described in Paper\,I is modified in two important aspects. 
\begin{itemize}
\item Instead of centring the windows on each point of a regular grid, the windows were centred on each observation. The estimate has minimum bias and variance at the centre, that is, at the observation, providing thus the best available estimate of the magnitude value corresponding to the primary pulsation.
\item The kernel length was adapted to the locally available information. In densely sampled periods, the inclusion of temporally distant observations may reduce variance, but increases bias at the centre, whereas in sparsely sampled periods, too narrow windows would have very high variance due to lack of data. Moreover, for long-period Cepheids, too short windows can contain too few full cycles for a reliable, low-variance estimate of the local period. The window halfwidth was therefore chosen individually for each observational epoch such that at least 120 residual degrees of freedom was left for the regression, and the window minimally included 15 full periods of the star (the window halfwidth is defined as the parameter $\sigma$ of the Gaussian kernel, and the kernel is truncated at  $\pm 3\sigma$).
\end{itemize}
 
This procedure resulted in window halfwidths varying between 50 and 250 days over the survey timespan. The window halfwidth fell below 100 days for example in the middle part of OGLE-II-III or in the first half of OGLE-IV, and reached occasionally 50 days within the OGLE-II-III phase and at the start of the OGLE-IV phase. 
In general, observations are more sparse in the second half of the OGLE-IV phase, increasing the necessary window lengths. As a result, the bias and the variance of the fitted magnitudes (and thus that of the residuals after pre-whitening) vary over the survey span. Despite of this, the adaptive window length offers the possibility for a more precise secondary mode detection than a constant kernel. A constant kernel width should be fixed long enough to produce reasonable estimates even in the most sparsely sampled survey periods. However, such a fixed-length window includes too many distant observations in densely sampled periods: we obtain a biased estimate at the central observation with a small variance and thus influential in the  secondary mode search. This bias is very hard to estimate, most notably because it depends on the true unknown process. Shortening the window reduces the bias but increases the variance, and therefore decreases the reliability and the influence of the central observation in the subsequent secondary mode search. However, this decreased reliability can be taken into account using the usual inverse error weighting, while the bias is practically impossible to account for. This motivated our decision to use the adaptive kernel lengths, and thus decrease the bias at the price of increasing somewhat the variance.

We refer to \citetalias{suvegesanderson17} for a detailed description of the kernel regression in the context of light curve analysis and a study of its performance on OGLE data.
The analysis was performed using the statistical software R \citep{R}.

\subsection{Input for the kernel method}
\label{subsec:input}

The local kernel method needs a few tuning parameters in order to restrict its scope of search, notably the (average) primary period and a reasonable estimate of the harmonic order describing the light curve. Assuming that the period of the primary pulsation does not deviate much from its average value, the nonlinear period search performed within the local kernel fit is initialised with the average period. Moreover, we suppose that we can use a single optimal local harmonic order of the light curves, the same through the whole survey span.

In order to find the best harmonic order for the local kernel fits, we performed a brief preliminary constant-model harmonic analysis of the light curves, separately to data from the two survey phases. Stars for which none of the phases recorded sufficient data (number of points less than 120 + number of coefficients in the model) were omitted from further analysis. Those with data only in one of the phases were kept, and treated identically to those observed throughout the whole survey period. Though the fitted models in the two phases were generally very similar, there were some cases  where the optimal model complexity was different in the phases. In all such cases, the more complex model was used in the kernel fits to avoid finding these omitted harmonics in the secondary mode search. We use the (non-weighted) mean of the two catalog values from OGLE-II/III and OGLE-IV as the initial value for the period in the local kernel fits as well as for the computation of period ratios. 
 
Systematic biases may arise from imperfections in the cross-calibration of survey phases, which can introduce e.g. constant offsets in magnitude due to changes in instrumentation. However, such offsets and the gaps between survey phases disturb much less the local model than the constant-parameter global model, because the local model uses data only from a small time interval. The lack of data in the gaps causes some bias at either end of the gap, but potential magnitude zero-point systematics between the phases influence the fits only in the few windows that span such gaps. In such instances, the downweighting by the Gaussian kernel and the inclusion of the polynomial term efficiently reduce the effect. There are in fact very few such windows: the average gap size between OGLE-II and OGLE-III is about 300 days, between OGLE-III and OGLE-IV, about 500 days, both of them longer than the average window size. Once the locally estimated magnitudes are subtracted from observed data, we can expect these residuals to have zero mean (by construction) and exempt from calibration systematics to a good approximation.

\subsection{Secondary period search and False Alarm Probabilities}
\label{subsec:fap}

By pre-whitening using the locally-fitted model magnitudes, we obtained the sequence of residuals and their estimated standard errors, spanning the whole OGLE II-III-IV survey length. For those stars that have data in both OGLE-II/III and OGLE-IV, the residual time series were  simply concatenated (cf. Section \ref{subsec:input}). 

We performed a generalised constant-parameter least-squares period search on these residuals \citep{zechmeisterkurster09} up to $8\, c/d$. For this first investigation of the full available population of Cepheid secondary modes in the Magellanic Clouds, we did not attempt to fit local kernel models for the secondary modes: all secondary modes were assumed to be perfectly stable, although this may be a bad assumption for many, possibly the large majority of them. For RR Lyrae stars, which show many similarities with Cepheids, temporal instabilities of the weak non-radial additional modes were indeed found \citep[e.g.][ using {\it Kepler} and {\it CoRoT} data]{moskaliketal15,szaboetal14}, which suggest that this might be the case with Cepheid secondary modes too.

We need a formal decision rule when to accept that a periodogram maximum indeed indicates a real oscillation mode. Within the classical statistical testing framework, this is equivalent to give its $p$-value, or its False Alarm Probability (FAP): the probability that a white noise sequence produces an equal or higher peak. The FAP is compared to a confidence level value $\alpha$, which is set in advance; if $\fap > \alpha$, we reject the existence of the mode. FAP for maxima of the degenerate, strongly dependent periodograms is a difficult problem, to which several good approximations based on extreme-value distributions  have been recently proposed in the literature \citep{baluev08, baluev13, suveges14, suvegesetal15}. Unfortunately, none of these is currently capable of dealing with general dependent errors in the time series, which is always the case for a secondary period search. Keeping in mind this caveat, we use the fastest of these methods \citep{baluev08} to compute FAP values. We adopt $\alpha = 0.05$ as the confidence level, aiming at completeness rather than purity and taking the results of simulations into consideration as well. Any mode with $\fap < \alpha$ is termed significant and will be called "modes". Modes where $0.0001 \leq \fap < 0.05$ are in fact "weak candidate modes", and it must be borne in mind that such peaks in the periodogram may be the product of noise with a relatively high probability. The  proportion of such weak candidate modes among all types of secondary modes analysed in this study is given in Table \ref{tab:secmodesByTypeAndFap}.  

\section{Simulations}
\label{sec:meth_simulations}

\subsection{Effects of OGLE sampling and errors}
\label{subsec:detectability}

\begin{figure*}
	\includegraphics[width=1.9\columnwidth]{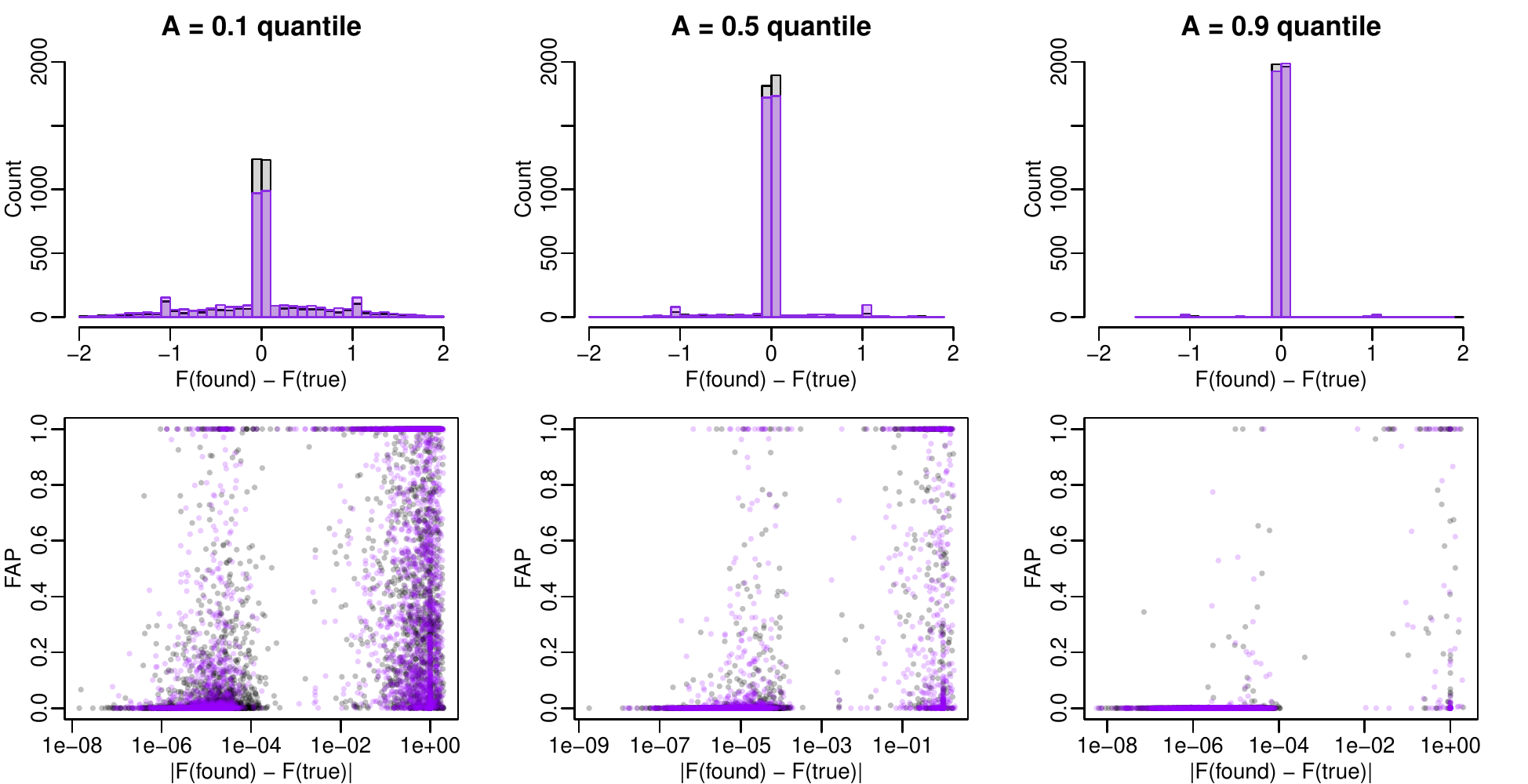}
\caption{Detection efficiency of weak modes using simulated mmag-level signals (left panels: small amplitude, equal to 1.2 mmag (the 0.1-quantile of all secondary amplitudes found to be significant at least at FAP level 0.05); middle panels: median amplitude, equal to 2.1 mmag; right panels: large amplitude corresponding to the 0.9-quantile, equal to 4.4 mmag). The top row shows the distribution of the differences between the true simulated frequency and the one found using LMC (grey) and SMC (violet) time samplings and corresponding error sequences. The bottom row shows the FAPs of the found signals as a function of the absolute value of the difference between the true simulated frequency and its estimate. Note the log scale of the $x$ axis in the bottom row.}
\label{fig:WeakSignalsSim}
\end{figure*}

Observational characteristics of time series from the two Magellanic Clouds are different in some important aspects. For instance, the time series in the SMC sample consist of on average more points than in the LMC (1232 against 917), while having a higher average error (0.0095 vs 0.0065\,mag). The interplay of these two and a few other factors such as the characteristics of the observing cadences induce different detection rates for weak periodic signals in the two Clouds, and bias the comparisons of their secondary mode populations. Thus, the recovery rates of weak signals of different amplitudes must be assessed via simulations. 

Using the fitted residual (secondary) models, we generated sinusoids with three different amplitudes corresponding to the 0.1, 0.5 and 0.9 quantile of all estimated secondary amplitudes (1.2, 2.1 and 4.4 mmag, resp.), with 1000 frequencies regularly spaced between 0.002 and 2 $c/d$. Each of the 3000 sequences was time-sampled eight times, four times using randomly selected OGLE-II-III-IV LMC cadences, four times using the SMC equivalent. Then noise was  added using a nonparametric bootstrap procedure (see \citetalias{suvegesanderson17}) to each simulated signal, according to the error bar sequence of the applied time sampling. We thus obtained four noisy signal sequences for each amplitude and frequency combination in each Cloud, i.e., a total of  4000 per Cloud and input amplitude. We then applied the same period search procedure to each simulated time series as for real OGLE data. 

Figure \ref{fig:WeakSignalsSim} presents the results. As the high peak at 0 in the top left panel shows, more than half of the weakest signals ($A = 1.2$ mag) give rise in the periodogram to a peak at approximately the correct frequency: about 2000 in the SMC, more than 2500 in the LMC. Taking a look at the lower panel nevertheless warns that a fraction of these frequencies have FAPs higher than 0.05, that is, they are deemed non-significant, and thus remain undetected. At this lowest amplitude, we can recover about half of the signals (having a significant FAP and $| F_{\mathrm{found}} - F_{\mathrm{true}}| < 0.00025$) in the LMC and 35\% in the SMC.

This detection efficiency improves with increasing amplitude: for the median amplitude (2.1 mmag), 10\% percent is missed in the LMC and 18\% in the SMC, whereas for the highest simulated amplitude (4.4 mmag),  2\% is missed in the LMC and 3\% in the SMC. A part of the missed signals are detected at either $F-1$ or $F+1$, the two first-order daily aliases, which for real Cepheids may be partially recovered using the structure of the characteristic frequencies in the $F_1-F_2$ plane. Weak signals are less efficiently detected in the SMC than in the LMC, the difference decreasing with increasing amplitudes.  Aliasing is also stronger in the SMC, presumably due to differences in time sampling. These facts imply that a larger fraction of a hypothetical weak secondary mode population may be missed in the SMC than in the LMC.

\subsection{Effects of pre-whitening with an adaptive kernel}
\label{subsec:simartefacts}

The simple picture above is complicated in our case by the likely existence of modulations in the primary period and harmonic parameters as well as by the time-dependent, adaptive kernel length pre-whitening.  

The performance of the adaptive-length kernel method has some natural limitations. First of all, the length of the kernel window puts an upper limit to the detectable modulation frequencies: shorter modulation periods than about half of the full window length cannot be detected. Since we need enough data to obtain a meaningful magnitude estimate at the window centre, we cannot decrease window length below a limit, which varies over the uneven time sampling of the survey. As a result, higher modulation frequencies are only partially or not at all modelled, causing an overdispersion in the data with respect to the photometric noise. 

To gain an idea about the effect of these limitations, we simulated primary pulsations with two different modulation periods, and added four different secondary modes to each. One of the simulated modulations had a frequency above the detection limit (0.025 $c/d$, corresponding to a period of 40 days, which cannot be precisely modelled under OGLE time sampling). The other simulation generated slow modulations which can be traced well in the present settings (0.00083 $c/d$, corresponding to a modulation period of 1200 days; see also \citetalias{suvegesanderson17}  for details on the performance of the kernel method). The amplitude chosen for the fast period modulation was 0.0007 $c/d$, based on what we have found in the intervals of the densest time sampling in the present study, and for the slow modulation, 0.00007 $c/d$ based on the results published in \citetalias{suvegesanderson17}. 
We have omitted amplitude modulations from the simulations, since (for Cepheids) they are typically  less perceptible than period modulations, and therefore are much less influential. We fixed the average primary frequency to be $F_1=0.55 c/d$ ($P_1 = 1.8182\, d$), and used the harmonic primary model of OGLE-LMC-CEP-0620, a typical single-mode FO Cepheid. The motivation to use an FO Cepheid as basic light curve was the fact that we found interesting, well-structured mode content in FO Cepheids, and a reliable analysis of their detectability and parameters requires a careful assessment of the effects of the instability of their primary pulsation and our non-standard pre-whitening technique.

Each of the two modulated primary pulsations were combined with each of four typical secondary modes found in the FO sample: a 0.62-mode ($0.62 P_1$), a 1.25-mode ($1.2 P_1$), a "far twin" ($0.97 P_1$), and a "close twin" ($0.994 P_1$). The amplitude of these modes was 1.6 mmag (approximately 80\% of the secondary modes found in our sample have higher estimated amplitudes). Each of the eight resulting signals were then time-sampled and noisified according to the observational cadence and errors of ten Cepheids (five of the LMC, five of the SMC, providing different window length patterns). Finally, these 80 simulated time series were subjected both to the local kernel pre-whitening and to a constant-model pre-whitening. Subsequently, we performed period search on the residuals of both procedures, resulting in two estimates of the true frequency $F_2$ of the secondary mode.

\begin{figure*}
	\includegraphics[width=1.8\columnwidth]{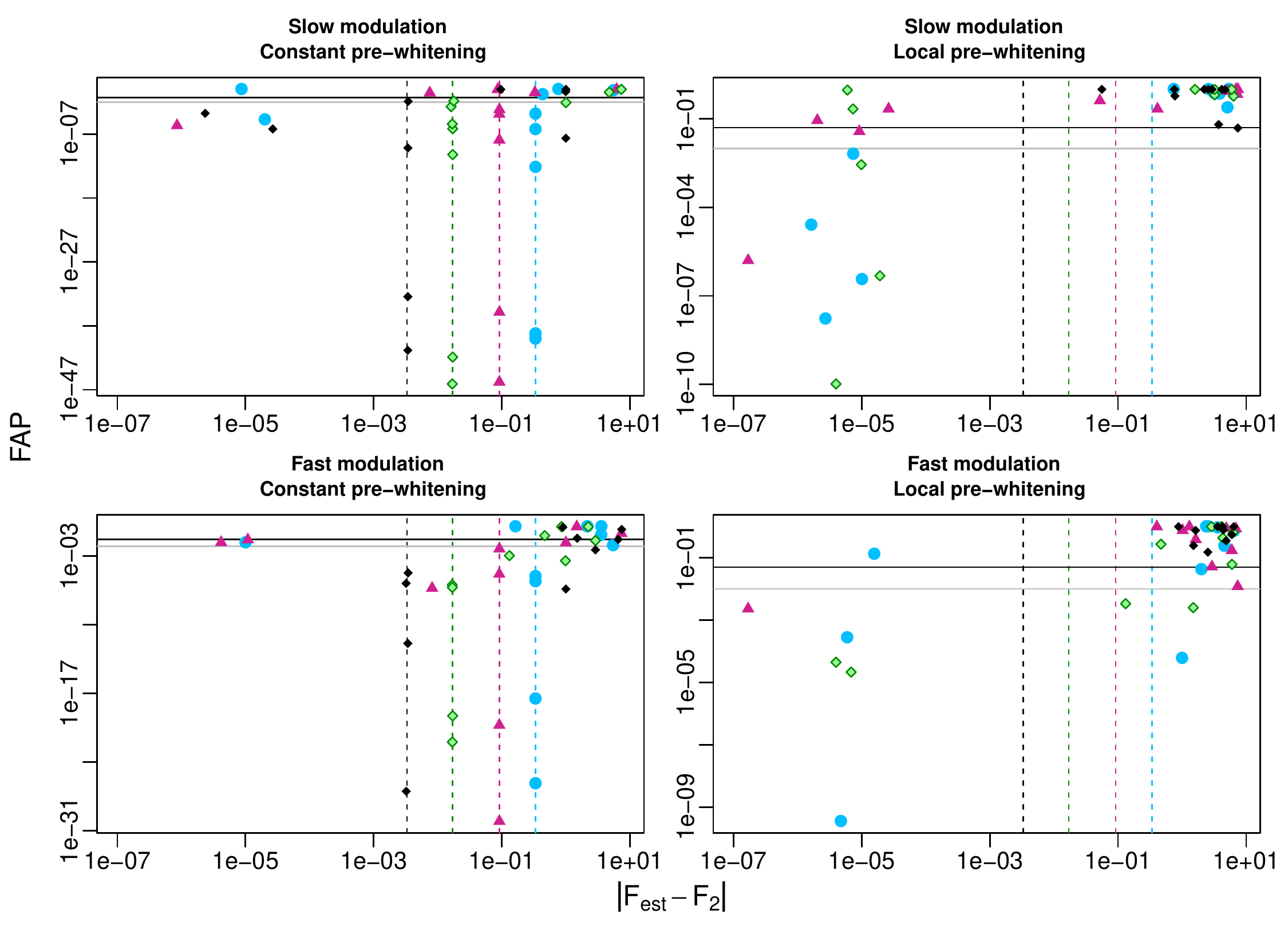}
\caption{FAP versus the bias of the secondary frequency estimate after constant pre-whitening (left column) versus local adaptive-kernel pre-whitening (right column), when the underlying modulation of the primary mode is slow (top row) or fast (bottom row). The coloured dots denote the four types of secondary modes simulated (black diamond: near twin, green diamond: far twin, purple triangle: 1.25-mode, blue dot: 0.62-mode). The coloured lines highlight the typical inaccuracy for each type of secondary mode, that is, the absolute difference of the average primary frequency from the true secondary frequency (when after the pre-whitening, a frequency very close to the average primary frequency is found again). The black horizontal line marks $\fap = 0.05$, the grey one, $\fap = 0.01$.}
\label{fig:local_vs_const_prew}
\end{figure*}

Figure \ref{fig:local_vs_const_prew} presents the FAP of the periodogram maxima versus the bias $| F_{\mathrm{est}} - F_{2}|$ of the estimates, the left column showing the results of constant-model prewhitening, the right, those of the kernel pre-whitening.

The best-estimated time series occupy the lower left quadrants of the panels: these show very small differences between the estimated and the true secondary frequency, with low (very significant) FAPs. The kernel pre-whitening (right column) places here markedly more objects than the constant pre-whitening (left column), implying that it is more efficient in the removal of the primary signal. This is so not only for slow modulations (top row), which can be well estimated by the kernel method, but also for fast modulations (bottom row), which can be estimated only with a rough approximation, and even that, only in the most densely sampled survey phases: the kernel method is able to at least partially model these variations. The constant-model pre-whitening results in fewer correct detections (four when the modulation of the primary pulsation is slow, and three when it is fast), whereas for the kernel pre-whitening, these numbers are nine and five, respectively.

There are some objects also at low estimation bias but with high FAP (upper left part of the panels, above the horizontal lines of FAP = 0.05 and 0.01): though the  maximum of the secondary periodogram fell near to the correct frequency, it was too small to pass the detection criterion (FAP $< 0.05$). The simulated secondary modes were very weak, so this loss is natural. 

The lower right part, to the right of $| F_{\mathrm{est}} - F_{2}| \approx 0.001$, and below the lines, contains false detections: the periodogram peaks at an incorrect frequency, with a high FAP.  The first feature to catch the eye is the pattern of detections ordered along vertical lines in the left panels. These lines are located at the absolute difference between the average primary frequency and the true secondary frequencies. The detections aligned on these lines therefore represent cases when after constant pre-whitening, the secondary period search detects a signal very near to the primary frequency: a typical mistake made by the constant pre-whitening when the primary pulsation is in fact modulated. A couple of more detections, beside the lines, fall on aliases of the primary frequency. Many of these false peaks are very strong, with FAPs below $10^{-7}$. 
 
For the kernel pre-whitening, this region is empty when the modulation is slow (top right panel), thanks to its ability to correctly follow the modulation of the primary. A fast modulation reduces the performance: there are four incorrect detections with FAP $< 0.05$ in this quadrant in the lower right panel. However, three of these are aliases of the correct frequency. Since we can obtain a preliminary knowledge about the patterns of the $F_1 - F_2$ plane thanks to stronger non-aliased signals, these aliases may be recovered and the true mode identified, which still improves the anticipated performance of the kernel pre-whitening.

The local kernel pre-whitening fails on the close-twin simulations. Such secondary modes result in a slowly varying primary period and harmonic amplitudes, and therefore the time-dependent fits remove them from the light curve. The constant-model pre-whitening is more efficient in the detection of these modes: in two cases out of ten it was able to detect them, when combined with slow changes (top left panel). Local pre-whitening represents a trade-off: in order to better detect secondary frequencies that are comparatively distant from the primary, we adopt a more complex model for pre-whitening, thereby sacrificing the ability to detect very close-by secondary frequencies. As Figure \ref{fig:local_vs_const_prew} shows, the kernel settings allow us to recover frequencies at $0.97F_1$, with efficiency depending on both the time sampling and amplitude of the signal.

The local kernel pre-whitening thus improves on the commonly used constant-model pre-whitening whenever the primary oscillation is modulated, and the secondary frequencies of interest are reasonably distant from the primary frequency. It reduces drastically the number of falsely detected "twin peaks" (cf. \citetalias{suvegesanderson17}) due to the remnants of an insufficiently modelled primary mode. Additionally, despite not allowing for the detection of secondary modes very close to the primary, it increases the chances to find weak secondary modes outside a narrow range. We examine this narrow range more in detail in the next section.

\subsection{Secondary frequencies close to the primary frequency}
\label{subsec:simcloseby}

\begin{figure*}
	\includegraphics[width=2\columnwidth]{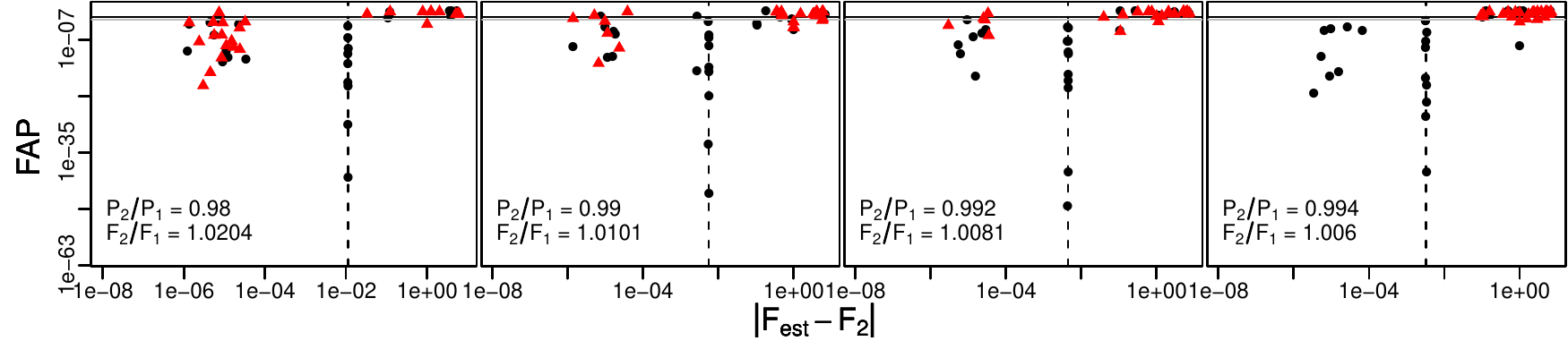}
\caption{FAP versus the bias of the secondary frequency estimate after constant and local pre-whitening for ten particular primary/secondary period ratios using simulated time series (the ratios $P_2/P_1$ and its inverse $F_2/F_1$ are shown in the bottom left corner of the panels). The coloured dots denote the two different pre-whitening techniques, black dots indicating the constant-parameter model, red triangles the local kernel model. The dashed vertical lines highlight the typical inaccuracy of the constant pre-whitening, that is, the absolute difference of the average primary frequency from the true secondary frequency (when after the pre-whitening, a frequency very close to the average primary frequency is found again). The black horizontal line marks $\fap = 0.05$, the grey one, $\fap = 0.01$.}
\label{fig:local_vs_const_prew_closebyfr}
\end{figure*}

The limit separation under which kernel pre-whitening removes secondary modes depends mainly on the kernel halfwidth. However, as a consequence of the adopted adaptive kernel length, this limit separation is greater in densely sampled parts of the time series and smaller in sparsely sampled periods. Depending on frequency separation and time sampling, a close-by frequency might not be fully removed from the light curve. We performed simulations to trace the interval around the primary in which secondary frequencies become undetectable using these tuning parameters. We used the same primary period and light curve parameters as in Section \ref{subsec:simartefacts}, and modulated its period with a frequency of 0.002 $c/d$ and amplitude 0.0001 c/d (detectable with our kernel parameter choices). We added to this modulated primary frequency a sequence of secondary modes, with period ratios $P_2/P_1 \in \{0.94,0.96,0.98,0.99,0.992,0.994,0.996,0.998, 0.999,0.9995 \}$, and sampled each according to 26 LMC and 26 SMC observational cadences. The cadences were chosen to represent well the variety of time samplings in both Clouds. Noise was added to the pure simulated light curves using the OGLE photometric error bars at the observational times. 

The sequence of period ratios where the local pre-whitening starts to remove near-primary secondary modes is shown in Fig.~\ref{fig:local_vs_const_prew_closebyfr}. The detections along the lines of typical  inaccuracy of the constant-model pre-whitening are clearly visible in the panels, with very low FAP values. However, as the secondary frequency moves towards the primary (from left to right), the constant pre-whitening starts to improve (five detections when $P_2/P_1 = 0.99$, six when $P_2/P_1 = 0.992$, and eight when $P_2/P_1 = 0.994$). Along this period ratio sequence, the timescale of the modulations caused by the secondary frequency of interest reaches the limits where the local kernel models are able to fit those modulations reasonably well, and thus, kernel pre-whitening removes the signal of interest too. The plots suggest that with the settings of this study, secondary signals with a period ratio below 0.98 can be detected with a good efficiency. Above this ratio, detection efficiency decreases sharply, and above  $P_2/P_1 = 0.994$, it drops practically to zero. 

This drop of detection efficiency motivates our decision to omit here the analysis of secondary modes close to the primary frequency. Though all such modes found in the analysed sample are included in the percentages given in the next section, it must be borne in mind that a number of close-by modes were not detected, and might further raise the fraction of secondary modes in Cepheids. Since these may well form a group of their own, and the local pre-whitening may partially or fully remove a part of them, the structure of the group and its general features cannot be reliably analysed.

\section{Results}
\label{sec:results}

\subsection{Overview of the secondary mode content}
\label{subsec:res_overview}


\begin{table*}
 \caption{Occurrence of secondary modes in the two Magellanic Clouds, ordered by significance and primary mode.}
 \label{tab:secmodesFap}
 \begin{tabular}{lrrrrr}
  \hline
   Cloud & Mode & $\mathrm{FAP} \! < \! 10^{-6}$ & $10^{-6} \! < \! \mathrm{FAP} \! < \! 10^{-3}$ & $10^{-3} \! < \! \mathrm{FAP} \! < \! 0.05$ & No sec.mode \\
  \hline 
  \multirow{4}{*}{SMC} & All modes & 255 (5.6\%) & 280 (6.1\%) & 1014 (22.3\%) & 3008 (66\%) \\
     & -- FU & 62 (2.3\%) & 104 (3.9\%) & 614 (22.8\%) & 1912 (71.0\%) \\
     & -- FO & 189 (10.6\%) & 176 (9.9\%) & 380 (21.4\%) & 1031 (58.1\%) \\
     & -- SO & 4 (4.5\%) & 0 (0.0\%) & 20 (22.5\%) & 65 (73.0\%) \\
  \hline 
  \multirow{4}{*}{LMC} & All modes & 504 (12.1\%) & 315 (7.6\%) & 596 (14.4\%) & 2734 (65.9\%) \\
     & -- FU & 66 (2.8\%) & 82 (3.4\%) & 351 (14.7\%) & 1882 (79.0\%) \\
     & -- FO & 434 (24.9\%) & 233 (13.4\%) & 242 (13.9\%) & 833 (47.8\%) \\
     & -- SO & 4 (15.4\%) & 0 (0.0\%) & 3 (11.5\%) & 19 (73.1\%) \\
  \hline 
 \end{tabular}
\end{table*}

Table \ref{tab:secmodesFap} provides a quantitative overview of the presence of secondary modes in the analysed Cepheid population in the Magellanic Clouds. 
\begin{itemize}
\item Stars with any secondary modes with $\fap < 0.05$ are present in the same fraction of stars in the LMC and the SMC (both about 34\%). However, the simulations presented in Section \ref{subsec:detectability} suggest that in the SMC, a higher fraction of weak secondary modes can go undetected, and thus, it is likely that in the SMC, there is a somewhat higher overall occurrence rate than in the LMC. 
\item Strong, highly significant modes are less frequent in the SMC than in the LMC (5.6\% against 12.1\%), and conversely, weak, barely significant modes are more frequent in the SMC than in the LMC (22.3\% against 14.4\%). The simulations (cf. Section \ref{subsec:detectability}) indicate that while the detection rate is very similar in the two Clouds for strong signals and is close to completeness, a significantly larger part of the weak modes is lost in the SMC than in the LMC. This increases further the likely population of weak candidate modes in the SMC. Thus, it seems likely that the Cepheids of the different Clouds are genuinely different in their secondary mode content. In Section \ref{subsec:secamp}, we give an overview about the distribution of the estimated first harmonic amplitude of the secondary modes.
\item The fraction of FU Cepheids with a secondary mode is higher in the SMC than in the LMC (29\% against 21\%). In both, most of them are very weak ($10^{-3} < \fap < 0.05$). Since weak modes are less efficiently detected in the SMC than in the LMC, it is likely that FU Cepheids in the SMC have secondary modes more frequently than the fraction we find, and therefore, the difference between the Clouds is even higher than our estimates.
\item The fraction of FO Cepheids with a significant secondary mode is lower in the SMC than in the LMC (41.9\% against 52.2\%); this difference might be in reality smaller (though probably not vanishing) if we take into account that in the SMC, a larger part of the weak modes is undetected. Among SMC FO stars, the weakly significant secondary modes are more frequent (21.4\% of all FO stars) than the very significant ones (10.6\% of all FO stars). The situation is converse in the LMC: there, 13.9\% have only a weakly significant secondary mode, and 24.9\% of all FO stars have a very significant one. Considering the detection efficiencies in the Clouds, the difference in the signal strength distribution between the LMC and the SMC is probably even greater.
\item The fraction of second overtone (SO) Cepheids with a significant secondary mode is similar to the proportion among FU Cepheids, and very low as compared to FO objects, in both Clouds. However, SO Cepheids are too few in the sample to assess population features.
\end{itemize}


In the next two sections, we discuss the secondary mode content in the FU and FO stars, and analyse the groups of modes among the FO stars one by one. We investigate the secondary modes using the  $F_1 - F_2$ plane instead of the Petersen diagram, since mode patterns, alias structures and mode ambiguities are clearer, easier to identify there than in the Petersen diagram. In the latter, parasite frequencies, trends and aliases appear as curved lines, and intersections of different aliased modes are hard to spot. In the  $F_1 - F_2$ plane, parasite frequencies, such as atmospheric effects with a one-day period, trends, and their aliases show up as horizontal lines. Modes in a constant or nearly-constant ratio with the primary period appear as approximately straight lines with a zero intercept and with slope equal to the constant frequency ratio. Moreover, the aliases of these lines form parallel straight lines shifted by $\pm1,\pm 2, \ldots c/d$, either with the same slope, or minus one times the slope. In the intersection of different lines, the mode is ambiguous. 

\subsection{Fundamental-mode Cepheids}

\begin{figure*}
	\includegraphics[width=1.98\columnwidth]{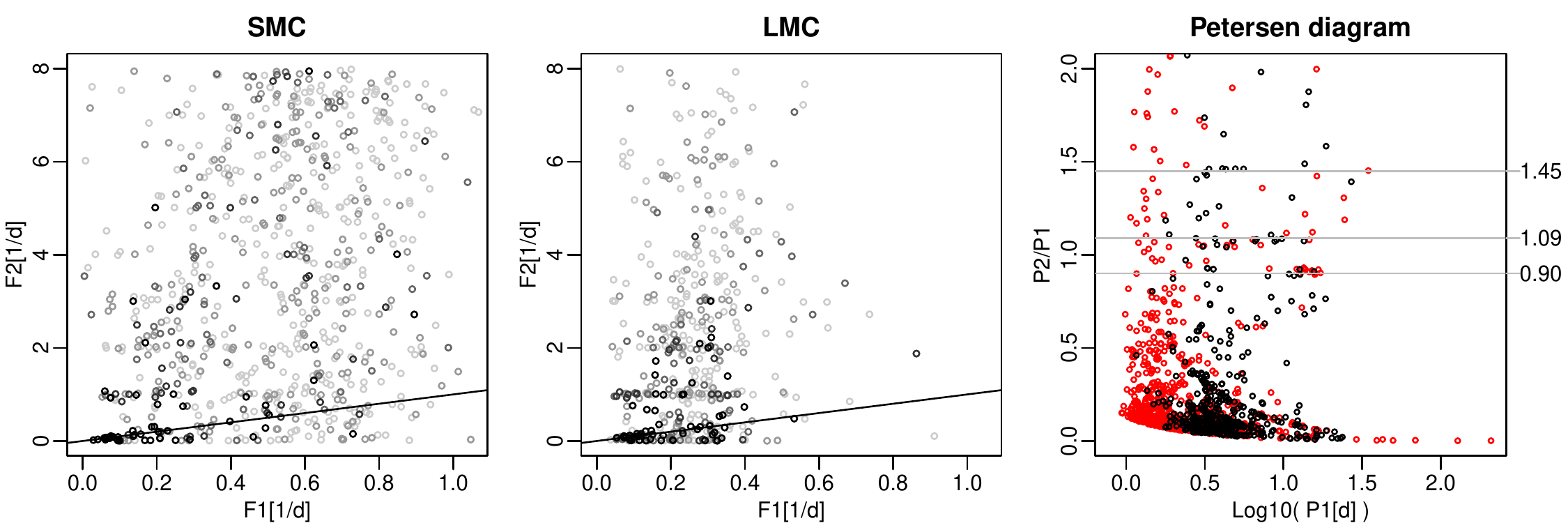}
\caption{The secondary frequencies versus the primary pulsation frequencies for fundamental Cepheids in the SMC (left panel) and in the LMC (middle panel), as well as the Petersen diagram (right panel). In the $F_1 - F_2$ planes, the shade of the colours indicate significance; light shades correspond to barely significant secondary frequencies, dark shades to strongly significant ones. The thin black line indicates $F_1 = F_2$. In the Petersen diagram, the red colour indicates SMC stars, the black, LMC ones. The grey lines highlight possible (dubious) local structures at frequency ratios of about 0.9, 1.09 and 1.45.}
\label{fig:f1_f2_fu}
\end{figure*}

\begin{figure}
\begin{center}
	\includegraphics[width=0.9\columnwidth]{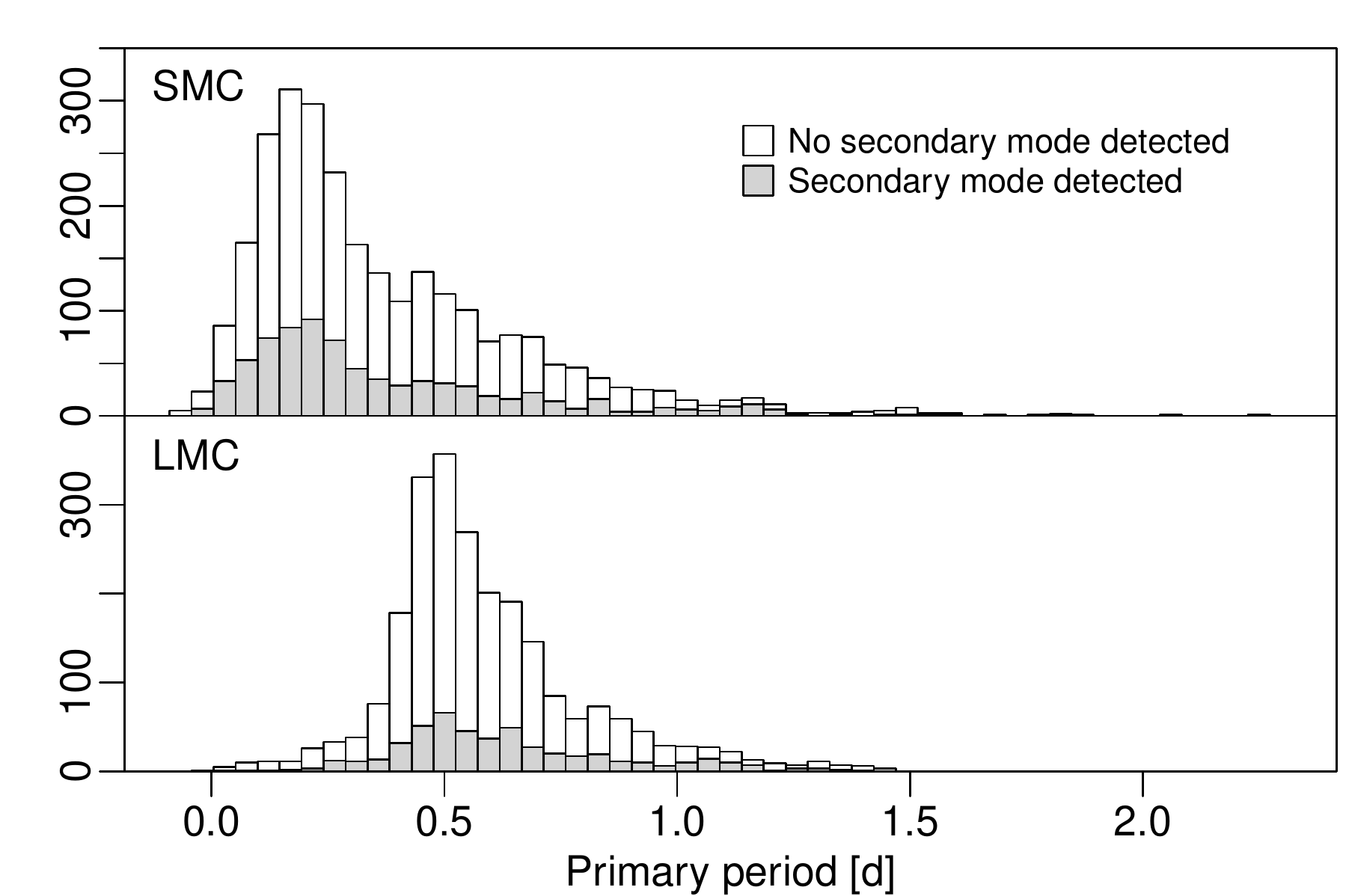}
\caption{Histogram of the primary periods of FU Cepheids in the sample. The fraction of FU Cepheids with a significant secondary mode is shown in grey. Top panel: SMC, bottom panel: LMC.}
\label{fig:hist_primperiodsFU}
\end{center}
\end{figure}

Figure \ref{fig:f1_f2_fu} visualises the secondary frequencies among the fundamental Cepheids on the $F_1 - F_2$ plane and in the Petersen diagram.  In the  $F_1 - F_2$ plane  no obvious pattern strikes the eye, apart from the horizontal lines formed by parasite frequencies of terrestrial origin, trends of the mean magnitude, and the $F_2 \approx F_1$ line or their aliases. There are a few short line-like sequences scattered over the plane, but these span only a very small fraction of the $F_1$ range. Most of the secondary modes outside these structures are relatively weak ($\fap > 10^{-4}$) in both Clouds, and are nearly uniformly distributed over the $F_1 - F_2$ plane. Figure \ref{fig:hist_primperiodsFU} shows that the distribution of the primary periods with and without significant secondary mode is very similar.

In the Petersen diagram (rightmost panel of Figure \ref{fig:f1_f2_fu}), the strongest features are the curved lines in the bottom third of the plot, corresponding to the (approximate) integer frequencies of the parasite signals and trends. We highlight three further weak structures at period ratios 0.9, 1.09 and 1.45. These features appear weak, composed only of a few stars. The strongest of these clusters is concentrated around $\log P_1 = 1$ and $P_2 / P_1 = 0.9$, with mostly very significant secondary signals in the SMC and mostly barely significant ones in the LMC. The loose, sparse band around $P_2 / P_1 = 1.09$ also contains some highly significant secondary modes (both in the LMC and in the SMC), along with some weakly significant ones. The third one, around  $P_2 / P_1 = 1.45$,  consists of only 5 LMC stars, and these are only marginally significant with one exception. The former two structures might be related to or affected by residuals of strong modulations in the primary pulsation, and therefore their statistical analysis can be unreliable (cf. Section \ref{subsec:simcloseby}), while the third group's existence is doubtful. The rarity and the weakness of these potential modes require large samples of long, temporally densely sampled time series of Cepheids for a secure detection. We thus do not pursue their analysis further in this study.

This unstructured, patternless Petersen diagram of FU Cepheids contrasts with findings in fundamental mode RR Lyrae stars (RRab), which show, in addition to the classical double-mode fundamental-first overtone stars, a host of other weak modes. \citet{molnaretal17} gives a comprehensive overview of these based on mostly photometry from space missions ({\it Kepler, K2, MOST, CoRoT}).  \citet{benkoetal14} investigates the possible multi-periodicity of the Blazhko modulation in 15 {\it Kepler} RRab stars, revealing a weak excitation of the second overtone in three of them. \citet{smolecetal16b} and \citet{prudiletal17} present and discuss double radial-mode RR Lyrae (RRd) stars with unusual primary periods and unusual period ratios in the OGLE database. No similar features, groups of additional modes can be  identified in our sample of FU Cepheids, raising the question why the fundamental mode of these two classical pulsator types are so different, while their first overtone mode shows extensive similarities as we discuss in the next section.

\subsection{First overtone Cepheids}

\begin{figure*}
	\includegraphics[width=1.7\columnwidth]{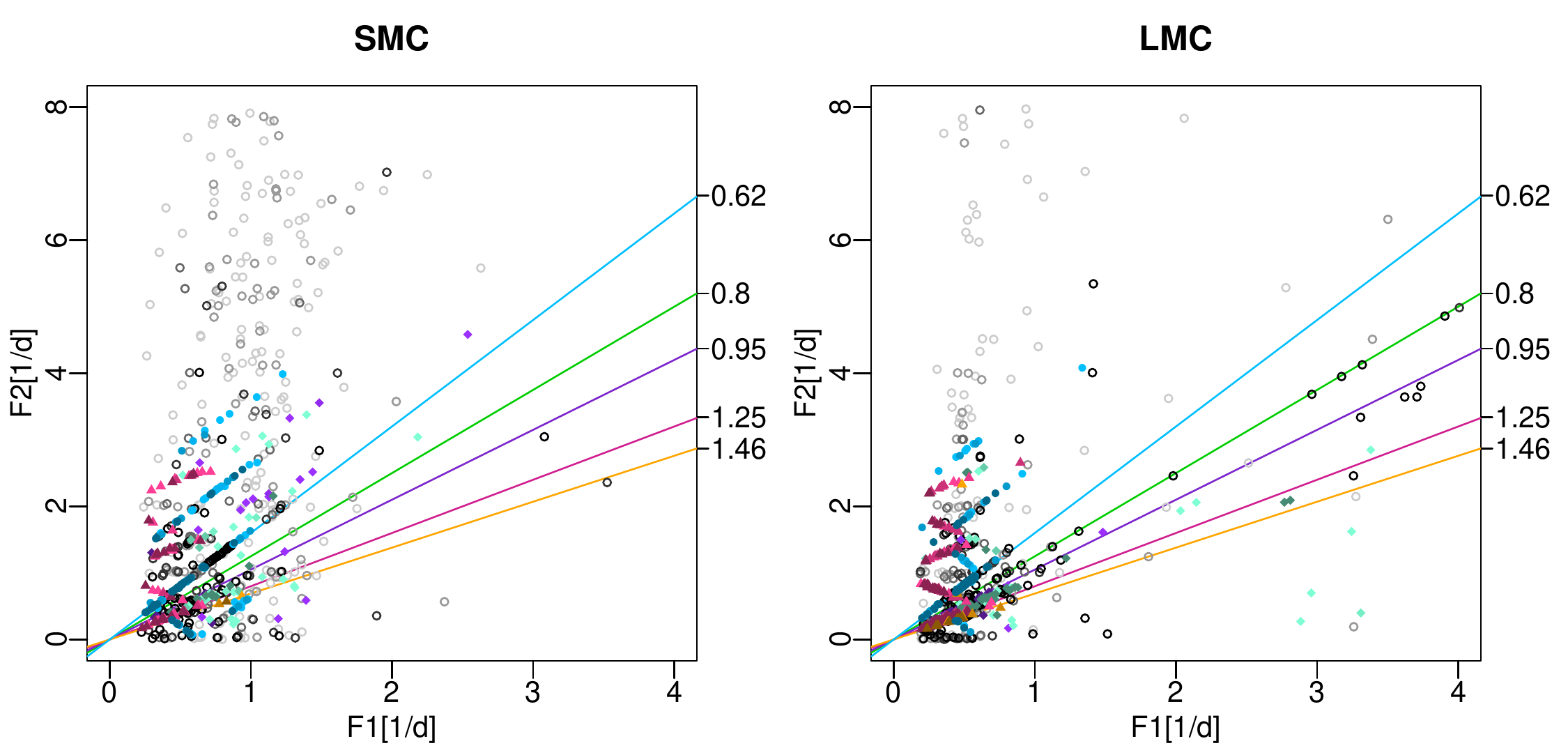}
	\includegraphics[width=1.7\columnwidth]{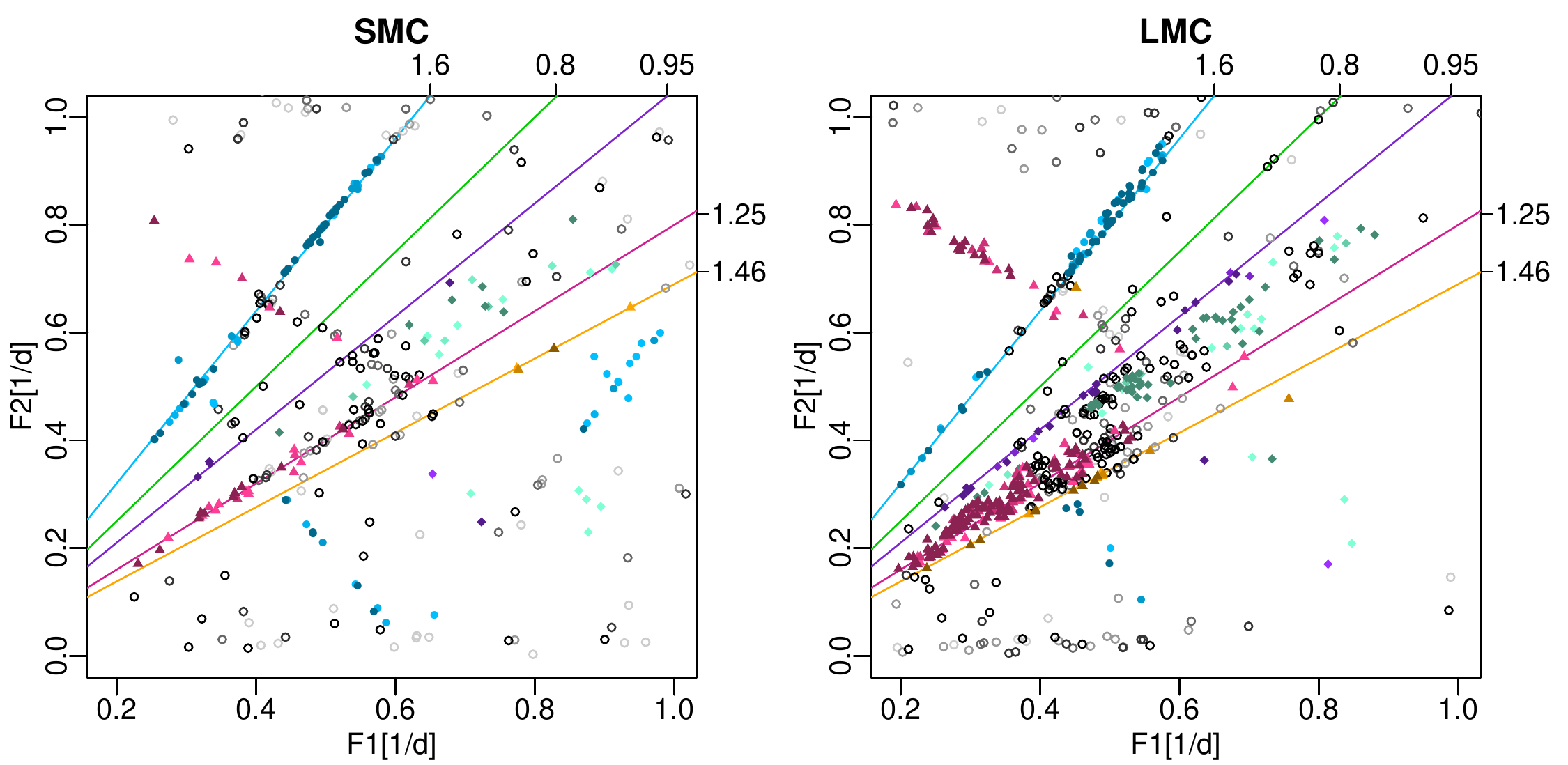}
\caption{The secondary frequencies versus the primary pulsation frequencies for first overtone Cepheids in the SMC (left column) and in the LMC (right column). The top row shows the complete $F_1-F_2$ plane. The second row is a zoom-in on the densest region. Purple triangles: potential 1.25-mode objects ($F_2 \approx 0.8 F_1$ and its aliases); orange triangles: potential 1.46-mode objects ($F_2 \approx 0.69 F_1$ and its aliases); blue dots: potential 0.62-mode objects ($F_2 \approx 1.6 F_1$ and its aliases); violet diamonds: potential 0.95-mode objects ($F_2 \approx 1.05 F_1$ and its aliases); green diamonds: a cluster of near-primary modes not in an $F_2 \approx \mathrm{constant}\times F_1$-type relationship with the primary frequency; empty grey circles: frequencies not belonging to either of the former clusters. The shade of the colours indicate significance; light shades correspond to barely significant secondary frequencies, dark shades to strongly significant ones. The green line indicates the location of second overtone frequencies. The numbers on the top and right edges of the panels at the end of lines refer to the corresponding mode (0.62, 0.95, 1.25 and 1.46 as defined in the text, 0.8 corresponding to the first/second overtone double mode).}
\label{fig:f1_f2_fo}
\end{figure*}

\begin{figure*}
\begin{center}
	\includegraphics[width=2\columnwidth]{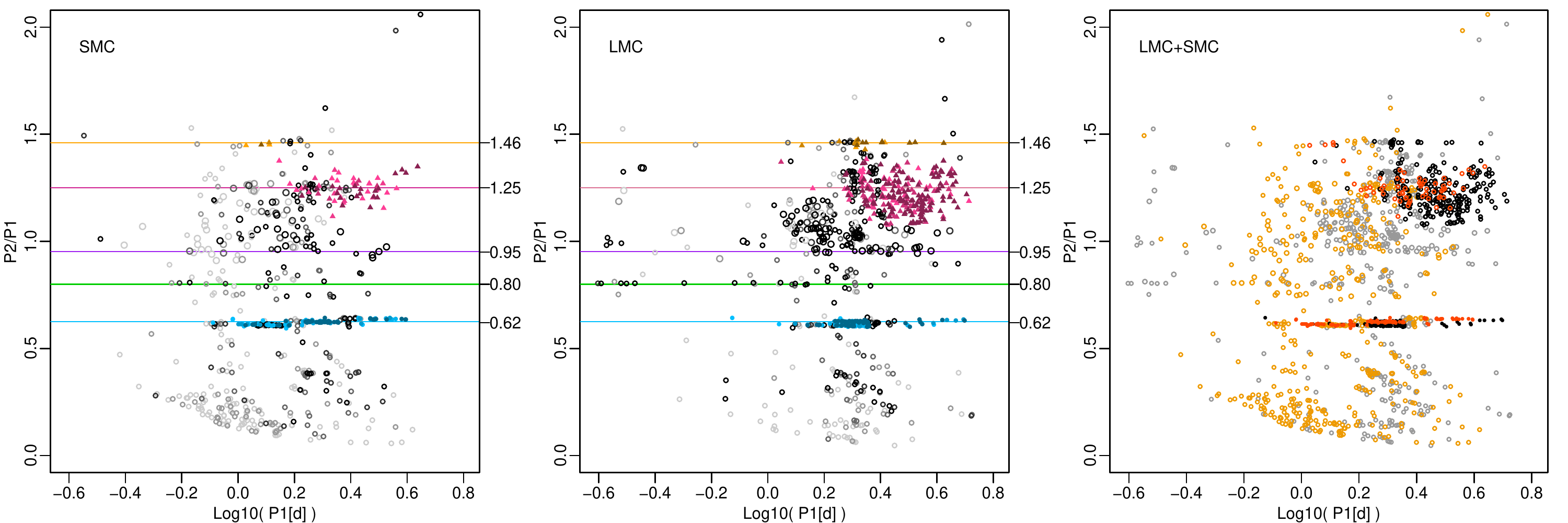}
\caption{ The Petersen diagram of the secondary modes for first overtone Cepheids. The leftmost plot refers to the SMC, the middle panel, to the LMC. The meaning of the colours and the shading in these panels is the same as in Figure \ref{fig:f1_f2_fo}. The numbers on the right side of the plots indicate the period ratios emphasised by the lines. The rightmost panel shows the same Petersen diagrams in the two Magellanic Clouds combined. Here, filled symbols highlight unambiguously identified 0.62- and  1.25-modes, while empty circles indicate everything else. Orange and red colours correspond to the SMC, grey and black to the LMC. Among the unambiguous identifications, we distinguish between aliased and non-aliased detections. Red and black filled symbols indicate non-aliased unambiguous cases, while orange and grey filled symbols aliased unambiguous ones. The plotted values of $P_2 / P_1$ are de-aliased (see text) for the unambiguously identified modes (that is, for the filled symbols), while they are the ratio of the secondary period with the highest periodogram peak to the primary period for the non-identified modes (empty orange and red circles).}
\label{fig:petersen_full}
\end{center}
\end{figure*}

\subsubsection{Mode identification and census}

\begin{table}
\begin{center}
 \caption{The definition of the secondary modes  of the FO Cepheids for the purposes of this study. The relatively complex expression for the 1.25-mode reflects the gap separating it from the cluster with a secondary frequency near the primary around $F_1 \in [0.5, 0.6]$.}
 \label{tab:defmodes}
 \begin{tabular}{lc}
  \hline
   Term & Definition \\
  \hline 
   0.62-mode  & $1.55F_1 < F_2 \leq 1.7F_1 $  \\
   0.95-mode  & $F_1 < F_2 \leq 1.1F_1$ \\
   1.25-mode  & $0.72F_1 < F_2 \leq \min\{0.14 + 0.6F_1, 0.93F_1\} $ \\
   1.46-mode  & $0.67F_1 < F_2 \leq 0.7F_1 $ \\
  \hline 
 \end{tabular}
\end{center}
\end{table}

\begin{table*}
 \caption{Occurrence of the different types of secondary modes in the FO Cepheids in the two Magellanic Clouds. The percentages are with respect to the total FO population in the Cloud.}
 \label{tab:secmodesByTypeAndFap}
 \begin{tabular}{llrrrrrrrr}
  \hline
  \multicolumn{2}{c}{Type} & \multicolumn{2}{c}{Total} & \multicolumn{2}{c}{$\mathrm{FAP} < 10^{-6}$} & \multicolumn{2}{c}{$10^{-6} < \mathrm{FAP} < 10^{-3}$} & \multicolumn{2}{c}{$10^{-3} < \mathrm{FAP} < 0.05$} \\
          & & SMC & LMC & SMC & LMC & SMC & LMC & SMC & LMC \\
  \hline 
  \hline 
  \multirow{2}{*}{0.62-mode} & Unambiguous & 139 (7.8\%) & 120 (6.9\%) & 50 (2.8\%) & 57 (3.3\%) & 41 (2.3\%) & 35 (2.0\%) & 48 (2.7\%) & 28 (1.6\%) \\
                            & All possible & 271 (15.3\%) & 209 (12.0\%) & 90 (5.1\%) & 105 (6.0\%) & 77 (4.3\%) & 59 (3.4\%) & 104 (5.9\%) & 45 (2.6\%) \\
  \hline 
  \multirow{2}{*}{1.25-mode} & Unambiguous & 65 (3.7\%) & 234 (13.4\%) & 23 (1.3\%) & 133 (7.6\%) & 16 (0.9\%) & 67 (3.8\%) & 26 (1.5\%) & 34 (2.0\%) \\
    & All possible & 153 (8.6\%) & 343 (19.7\%) & 54 (3.0\%) & 184 (10.6\%) & 50 (2.8\%) & 100 (5.7\%) & 49 (2.8\%) & 59 (3.4\%) \\
  \hline  
  \multirow{2}{*}{1.46-mode} & Unambiguous & 4 (0.2\%) & 19 (1.1\%) & 1 (0.1\%) & 9 (0.5\%) & 1 (0.1\%) & 7 (0.4\%) & 2 (0.1\%) & 3 (0.2\%) \\
    & All possible & 30 (1.7\%) & 37 (2.1\%) & 7 (0.4\%) & 16 (0.9\%) & 9 (0.5\%) & 13 (0.7\%) & 14 (0.8\%) & 8 (0.5\%) \\
  \hline  
  \multirow{2}{*}{SO} & Unambiguous & 32 (1.8\%) & 16 (0.9\%) & 7 (0.4\%) & 8 (0.5\%) & 6 (0.3\%) & 3 (0.2\%) & 19 (1.1\%) & 5 (0.3\%) \\
   & All possible & 78 (4.4\%) & 112 (6.4\%) & 18 (1.0\%) & 52 (3.0\%) & 24 (1.4\%) & 33 (1.9\%) & 36 (2.0\%) & 27 (1.5\%) \\
  \hline 
  \multirow{2}{*}{Near-primary} & Unambiguous & 70 (3.9\%) & 123 (7.1\%) & 15 (0.8\%) & 76 (4.4\%) & 14 (0.8\%) & 20 (1.1\%) & 41 (2.3\%) & 27 (1.5\%) \\
   & All possible & 227 (12.8\%) & 250 (14.4\%) & 67 (3.8\%) & 144 (8.3\%) & 53 (3.0\%) & 43 (2.5\%) & 107 (6\%) & 63 (3.6\%) \\
  \hline 
  \multicolumn{2}{l}{Other (neither of the above)} & 176 (9.9\%) & 142 (8.2\%) & 13 (0.7\%) & 26 (1.5\%) & 25 (1.4\%) & 35 (2.0\%) & 138 (7.8\%) & 81 (4.6\%) \\
  \hline 
 \end{tabular}
\end{table*}

The $F_1 - F_2$ plane for the first overtone Cepheids in Figure \ref{fig:f1_f2_fo} exhibits a rich structure, providing clear, insightful information about the secondary mode content in the FO Cepheids. In addition to the horizontal lines of parasite frequencies, trends and their aliases, we can immediately identify four different types of secondary modes. 

The first is the recently identified non-radial mode, where $P_2 / P_1 \in (0.6, 0.65)$ \citep{moskalikkolaczkowski09, soszynskietal10, soszynskietal15a, smolecsniegowska16}. In Figure \ref{fig:f1_f2_fo}, this mode appears as parallel lines (in blue) with a slope of about $\pm1.62 \approx 1/0.63$. The aliases can be clearly identified up to $| 1.62 F_1 \pm 2 c/d |$. We identify a second, new candidate secondary mode that is aligned along $F_2 \approx 0.8 F_1$, i.e., where the period ratio is approximately 1.25 (in purple); this sequence is relatively clear-cut in the SMC, and forms a densely populated, broad, slightly curved stripe in the LMC. A third new candidate mode is situated along the line $F_2 \approx 0.685 F_1$, that is, at period ratios around $1.46$ (in orange). 

We further identify a cluster of modes occurring in majority with primary periods between 1.7 and 2 days (that is, $F_1 \in [0.5, 0.6]$) at secondary frequencies close to the primary ($F2 \in [0.9F_1, 1.06F_1]$). However, this cluster does not follow a tight  linear relationship of the form $F_2 \approx c F_1$. It is not clear how to define it, specifically whether to include an apparent, very loose line-like continuation. Moreover, part of this cluster can be missing: very close-by beating modes have been fully removed by the kernel pre-whitening. Others, farther away from the primary frequency, though detected, can be nevertheless affected: the signal might be partially or completely removed in densely sampled time intervals, while it can be detected in sparsely sampled time intervals (see Section \ref{subsec:simcloseby}). Since the results are likely to be unreliable, we do not analyse further here this group. A potential fourth candidate mode can be one with $F_2 = 1.05F_1$ (indicated in violet in Figure \ref{fig:f1_f2_fo}). This sequence, which we tentatively refer to as the 0.95-mode, is quite clear in the LMC, but contains only very few stars in the SMC. The line forms the high-frequency boundary of the cluster of secondaries close to the primary, just above the stripe where local pre-whitening may remove secondary modes, and therefore may or may not be a subset of the near-primary group. Because of the same arguments as above, we also defer the analysis of this group to a study using different methods.

In addition to these candidate non-radial modes, we identified several secondary oscillations that are most likely weak second radial overtones; the majority of which were found in the LMC (green line in Figure \ref{fig:f1_f2_fo}).

The corresponding Petersen diagram can be seen in Figure \ref{fig:petersen_full}. The left and middle panels  show the modes separately for the LMC and the SMC, revealing a very similar structure of clusters and lines in the two Clouds. However, the clusters occur with different frequency in the SMC and in the LMC, and in addition, the  smaller average statistical significance of the modes SMC is clearly observable (by the lighter colours). The joined Petersen diagram of both Clouds is shown in the right panel, exhibiting clearly the lines of the 0.62- and the 1.46-mode, the band of the 1.25-mode, the cluster of the near-primary modes, and weakly hinting at the aforementioned candidate 0.95-mode. 

For the purposes of this study, we set up working definitions of the above mentioned secondary modes, which are summarised in Table \ref{tab:defmodes}. Moreover, since well-established terminology does not yet exist for such periodicities in Cepheids, we shall term the oscillations according to their approximate period ratio for the ease of discussion. Further studies and theoretical investigations are needed to clearly define these modes, and delimit their different types from each other according to their physical origin.

Table \ref{tab:secmodesByTypeAndFap} gives an account of the population sizes and relative frequencies of the different modes, separately by Cloud and FAP level. Since alias lines up to the second can be clearly identified in the plot for both the 0.62- and the 1.25-mode, we call "candidate $x$-mode" any periodicity with $\fap < 0.05$ that lies  either on the
$F_2 \approx F_1/x$ line, or the alias lines $F_2 \approx | \pm F_1/x \pm 1,2 |$ (where $x$ is the commonly used period ratio). "Unambiguous candidate $x$-modes" are those candidate  $x$-modes which lie on a single, uniquely identifiable line, or in the intersection of two different alias lines of the same mode, but not in the intersection of two or more lines belonging to two or more different modes. We did not include in the table Cepheids that were detected as blended or with an additional eclipsing variability, or had uncertain classifications.

We publish the list of all FO Cepheids that we found to have significant secondary modes in four online tables. Data from the LMC and the SMC are presented in two separate tables, and we provide both tables in two formats: one is a human readable file of which a few lines are shown in Table \ref{tab:allsecmodesSMC}, the other is a machine-readable semicolon-separated ascii file. The stars are ordered first according to their mode (0.62-mode, 1.25-mode, 1.46-mode, 0.95-mode, near-primary cluster, parasitic frequency (close to integer frequency), possible SO, possible FU, and Unknown), and within mode, according to their identifier. The list for the SMC indicates whether the star is in the list of \citet{smolecsniegowska16}.

\begin{landscape} 
\begin{table}
\caption{Online list of significant secondary modes in the Magellanic Clouds. The example shows parts of the table containing the SMC objects. The columns are as follows: 
$\,^a$: the identifier xxxx from the OGLE catalogue name OGLE-SMC-CEP-xxxx; $\,^b$: OGLE-III catalogue period (in days). $\,^c$: OGLE-IV catalogue period (in days). $\,^d$: Comment in the OGLE-III catalogue.  $\,^e$: Comment in the OGLE-IV catalogue.   $\,^f$: Secondary period found in this study (in days). If the mode belongs to a unique mode and alias sequence, this period is unambiguous, and (if necessary) can be de-aliased. In this case the de-aliased value is given in the table, and is typeset in black. If the mode lies in the intersection of two or more alias sequences, the given number is the period corresponding to the peak of the periodogram, typeset in blue.  $\,^g$: False Alarm Probability (cf. Section \ref{subsec:fap}).  $\,^h$: Secondary mode of the star. The mode is Unknown whenever the frequency lies in the intersection of alias sequences belonging to two different modes,  if it does not belong to any recognisable sequence, or if it belongs to a unique alias sequence of order 3 or higher. The mode is not Unknown only when the secondary frequency lies on a unique line, corresponding to the definitions given in Table \ref{tab:defmodes}, or a $\pm 1$ or $\pm 2$ alias thereof. $\,^i$: The possible modes for the secondary. The numbers in brackets indicate the order of the alias. For the SMC, S\&S16 indicates that the star is listed in \citet{smolecsniegowska16} as a 0.62-mode star.}
 \label{tab:allsecmodesSMC}
 \resizebox{1.35\textheight}{!}{  
  \begin{tabular}{rrrrrrrrr}
  \hline
  Identifier$^{a}$ & $P_1[d]$ OGLE-III$^{b}$  & $P_1[d]$ OGLE-IV$^{c}$ & Remarks OGLE-III$^{d}$  &    Remarks OGLE-IV$^{e}$ & $P_2[d]^{f}$  & FAP$^{g}$ & Mode$^{h}$ &  Possible modes (alias)$^{i}$ \tabularnewline
  \hline
0052 & 1.718381 & 1.718387 &  &  & 1.07203 & 2.56e-03 & 0.62-mode & 0.62-mode(1) \tabularnewline 
  0099 & 0.958971 & 0.958978 &  &  & 0.609492 & 1.34e-03 & 0.62-mode & 0.62-mode(2) \tabularnewline 
  0212 & 1.741011 & 1.741042 & secondary period of 1.08768 days & secondary period of 1.087820 d & 1.087555 & 9.74e-09 & 0.62-mode & 0.62-mode(0), S\&S16 \tabularnewline 
 ... & & & & & & & & \tabularnewline 
  0410 & 1.613567 & 1.613579 &  &  & 	\textcolor{blue}{1.987523} & 1.05e-05 & 1.25-mode & 1.25-mode(0), 1.25-mode(1) \tabularnewline 
  0422 & 2.884187 & 2.885381 & high rate of period change &  & 3.552812 & 2.68e-03 & 1.25-mode & 1.25-mode(0) \tabularnewline 
  0477 & 3.940331 & 3.940289 &  &  & 5.199118 & 3.42e-11 & 1.25-mode & 1.25-mode(1) \tabularnewline 
 ... & & & & & & & & \tabularnewline 
  0020 & 0.875214 & 0.875255 &  &  & 	\textcolor{blue}{0.19309} & 2.39e-03 & Unknown & 0.95-mode(4) \tabularnewline 
  0056 & 0.986021 & 0.985973 & secondary period of 0.60372 days &  & 	\textcolor{blue}{0.19574} & 2.29e-03 & Unknown & 0.95-mode(4), S\&S16 \tabularnewline 
  0064 & 2.151393 & 2.151426 &  &  & 	\textcolor{blue}{3.267815} & 3.87e-02 & Unknown &  \tabularnewline 
 ... & & & & & & & &
 \end{tabular}  
 }
\end{table}
\end{landscape}

\subsubsection{The 0.62-mode}
\label{subsubsec:0.6mode}

\begin{figure}
\begin{center}
	\includegraphics[width=0.85\columnwidth]{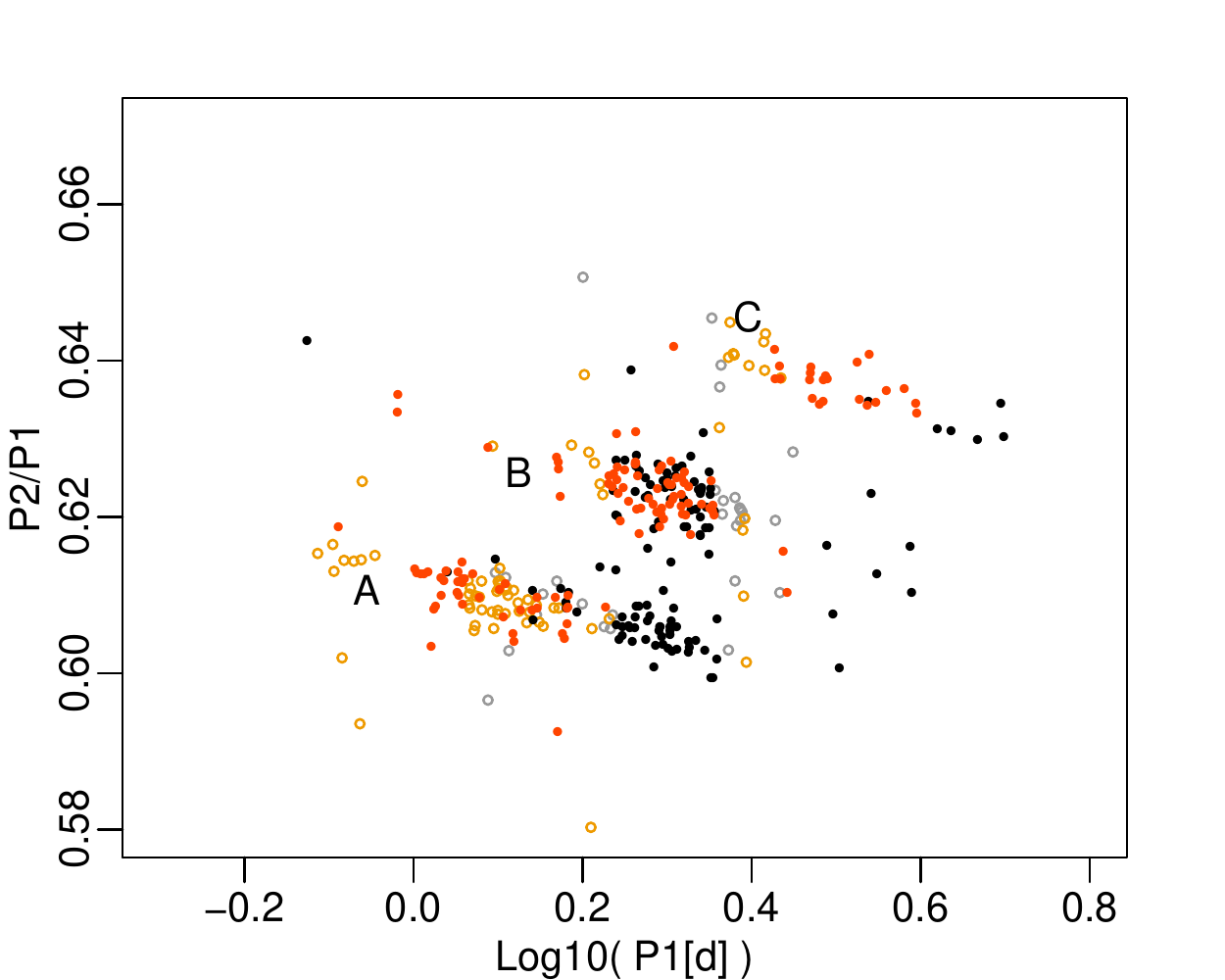}
\caption{Enlarged Petersen diagram in the region of the 0.62-mode, its subsequences denoted by A, B, and C. The colour code is the same as in the rightmost panel of Figure \ref{fig:petersen_full}. The plotted values of $P_2 / P_1$ are de-aliased (see text) for the unambiguously identified modes (i.e., filled dots), while they are the ratio of the secondary period corresponding to the highest periodogram peak and the primary period for the non-identified modes (empty orange and grey circles).}
\label{fig:petersen_06-mode}
\end{center}
\end{figure}

\begin{figure*}
	\includegraphics[width=2.1\columnwidth]{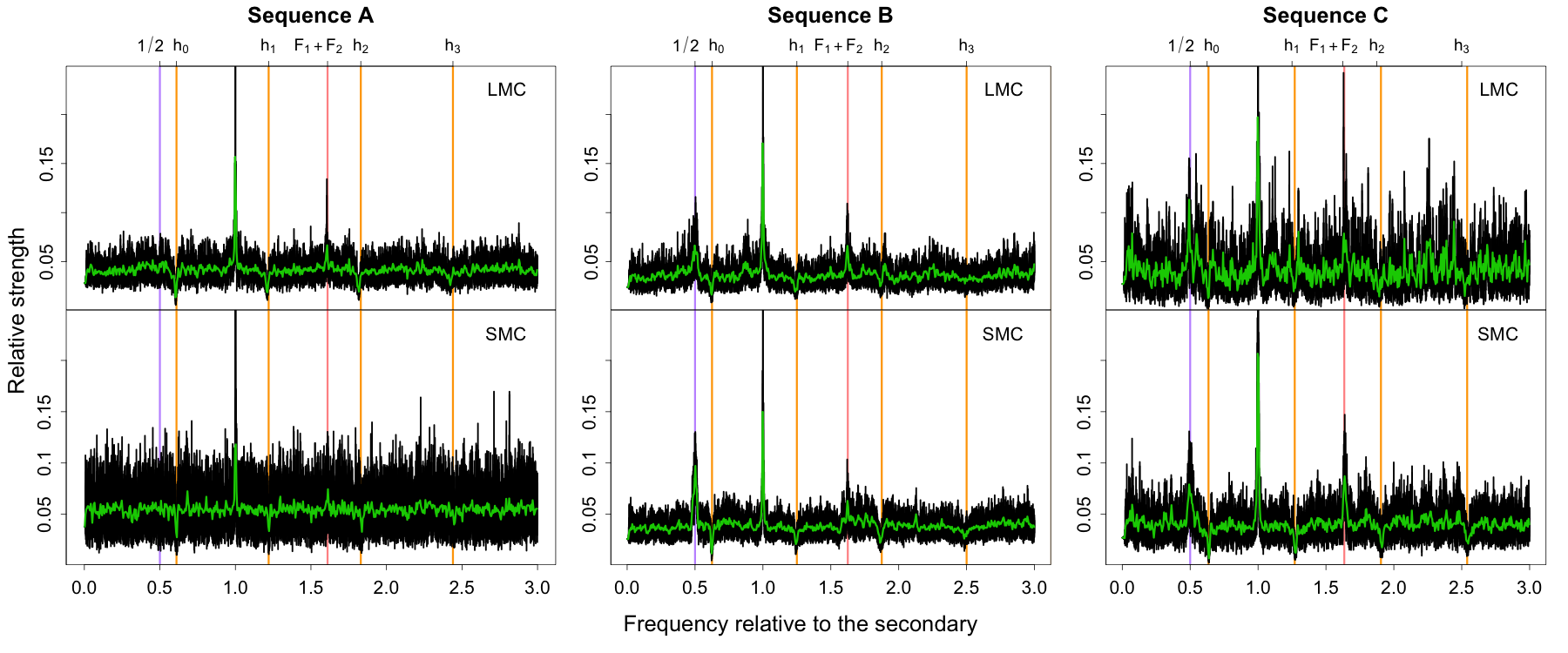}
\caption{Standardised average secondary periodograms, the average taken over those FO stars that have a secondary unambiguous non-aliased 0.62-mode  in the LMC (top panel), and in the SMC (bottom panel); see text for detailed explanation about the standardisation. A sliding average over 500 frequencies of the spectra is added to both (green). The vertical violet line denotes the position of the 1.25-mode (the 1/2 subharmonic, indicated by 1/2 at the top of the line), the orange lines indicate the locations of the primary frequency and its first three harmonics (denoted by $h_0, \ldots, h_3$ at the top), and the red line highlights another frequency around which the secondary spectra of the stars with a 0.62-mode tends to contain above-average power, likely the combination $F_1 + F_2$ (denoted by $F_1 + F_2$ at the top).}
\label{fig:oglenonrad_avgSecPgram}
\end{figure*}

The number of FO Cepheids exhibiting this non-radial secondary mode has been estimated to be 82 in the LMC and 127 in the SMC by \citet{soszynskietal15b}. The more precise predicted magnitudes of the primary light curve, given by the adaptive-length kernel pre-whitening technique, and the use of a good approximate FAP allow us to re-estimate these numbers at a total of 139 unambiguous candidates in the SMC and 120 in the LMC if considering only the unambiguous detections with $\fap < 0.05$. Counting all ambiguous candidates (lying in the intersection of the 0.62-mode or its first or second alias with some other mode or alias line in Figure \ref{fig:f1_f2_fo}) as 0.62-mode increases these numbers to 271 in the SMC and to 209 in the LMC. These latter numbers can be considered as upper limits of the fraction of such secondary modes that can be detectable with the adaptive-length kernel method on joined OGLE-II-III-IV data. 

The Petersen diagram, presented in Figure~\ref{fig:petersen_06-mode}, shows clearly the three sub-sequences of this mode. For the SMC, we recover the shift of the sequences: from A to C, they are concentrated at increasingly long primary periods. For the LMC, so far less frequently analysed in the literature because of the comparative scarcity of 0.62-mode stars there, we find a different pattern. The majority of the two lower sequences, A and B, occur at nearly the same primary periods in a narrow interval, although there are a few stars extending both sequences (A towards shorter periods, B towards longer ones, in agreement with the patterns in the SMC). There are also a few unambiguously detected objects belonging to sequence C in the LMC. All but one lie in the extension of the SMC sequence C towards longer periods; the only object at shorter periods is  in the middle of the SMC sequence C. In addition, there are a few more such candidates with an ambiguous mode identification. 

The origin of this secondary pulsation is most likely the excitation of a non-radial mode, although for a long time, there was no strong candidate as to which this mode is. Recently, \citet{dziembowski16} proposed that the 0.62-mode is in fact the first harmonic of a low-order non-radial, $l = 7, 8, 9$ mode, which becomes easier to observe than the mode's fundamental frequency due to two effects: first, the fundamental is geometrically nearly cancelled over the star's visible surface, and second, the first harmonic  is enhanced by nonlinear effects. Among the predictions of this hypothesis, one is that the fundamental may nevertheless appear in the spectrum of especially even-order modes, that is, in sequence B ($l = 8$). \citet{smolecsniegowska16} found, confirming this prediction, that the 1/2 subharmonic is present in about 35\% of SMC 0.62-Cepheids, in majority in sequence B, but also occurring in sequence C. Using this enlarged sample, we can investigate this question also in the LMC. 

The secondary periodogram of Cepheids showing the 0.62-mode is often very complex, as was also shown by \citet{smolecsniegowska16}. Searching for frequency ranges where the secondary periodograms tend to contain on average more power, we consider the standarised periodograms $Z^*(\nu) = Z(f) / \max Z(f)$ as a function of the frequency ratio $\nu = f / F_2$, instead of the usual periodogram. The standardisation leads to a common value of periodogram maxima: $\max\{Z^*(\nu)\} = 1$ corresponding to the peak of the secondary periodogram, located at $\nu = 1$ (corresponding to $F_2$) for all objects. Then we selected those Cepheids that have an unambiguous non-aliased 0.6 mode, and took the average of all such standardised  spectra (Fig.~\ref{fig:oglenonrad_avgSecPgram}), separately for each subsequence (A, B and C). If another mode with a typical frequency ratio is often present in the spectra, this average spectrum should show increased power at this ratio. 

Fig.~\ref{fig:oglenonrad_avgSecPgram} demonstrates that the 1/2 subharmonic is indeed present in sequences B and C in the SMC, and absent or much less prevalent in its sequence A, as \citet{smolecsniegowska16} observed. Based on the figure, we can extend this statement to LMC 0.62-mode stars. However, we find the subharmonic to be almost as strong in sequence C ($l=9$) as in sequence B ($l=8$) in the SMC, and in the case of LMC, maybe even stronger (although sequence C comprises only seven non-aliased modes in the LMC, so their average periodogram has a much larger scatter than the others). This is somewhat different from the findings of \citet{smolecsniegowska16} as well as from the theory of \citet{dziembowski16}, which predicts a stronger presence of this subharmonic in even-$l$ modes than in odd-$l$ modes.

Another, alternative explanation for the 1/2 subharmonic in the spectra of these stars could be period doubling, of which one indication is the presence of half-integer harmonics in the spectrum. The phenomenon of period doubling is assumed to be the first step towards a chaotic pulsation \citep{buchlerkovacs87, kovacsbuchler88, moskalikbuchler90, moskalikbuchler91}.  However, the folded light curves of a period-doubling star should also show alternating peak-to-peak amplitudes between consecutive pulsation cycles. Although the low amplitudes of the secondary pulsation and the noise makes it hard to ascertain, there is no indication of such alternating amplitudes in the data. In addition, the independent, frequent presence of another secondary mode, the 1.25-mode (see later in section \ref{subsubsec:1.2mode}) at the half-frequency of the 0.62-mode supports rather the interpretation of the subharmonic as the weakly present fundamental frequency of the mode. The half-frequency in the framework of period doubling is not expected to exist independently as a mode. If the subharmonic in the spectra of the 0.62-mode stars is due to period doubling,  the 1.25-mode should be explained in a different way, and  raises new questions for theory. 

Another peak in the average periodograms, at approximately the linear combination $F_1 + F_2$, is obvious in all subsequences, and is clearly dominant in sequences A and C ($l = 7$ and 9) over every other feature, the half-frequency subharmonic included. The dips at $F_1$ and its first few harmonics are due to the effect of pre-whitening. When minimising the sum of the squared errors of the fit, we fit also any random oscillatory component at this frequency that the noise happens to contain, and remove it from the time series, leaving less power in the periodogram at this frequency than would be expected for white noise. No other strong characteristic feature can be distinguished, although the (pattern-seeking) human eye is tempted to see a few candidate peaks (e.g., in sequence C in the LMC a bit above $\nu = 1.5$). 

Using the joined OGLE-II-III-IV data, we detect a 0.62-mode (or its 1- or 2-alias) in 98 of the SMC 0.62-mode stars listed by \citet{smolecsniegowska16}. Of these, 0.62-mode is the unambiguous dominant mode in only 49 cases, while in the 49 other stars, the frequency can admit an alternative explanation as well. A further 25 stars show other significant secondary frequencies: six have a dominant 1.25-mode, two may be explained as a second overtone, two fall close to a parasitic frequency, and fifteen more are ambiguous without a low-aliased 0.62-mode among the candidates. The 15 remaining stars in their list do not have a significant secondary mode in our study (OGLE-SMC-CEP-0708, 0833, 0841, 1053, 1516, 1976, 2528, 2593, 2597, 2910, 3143, 3249, 3624, 3903 and 4157), though the non-significant maximum of the secondary periodogram is at the 0.62-mode for 7 of them (1053, 1516, 2528, 2593, 2597, 2910, 3624). The discrepancies between this study and that of \citet{smolecsniegowska16} may be due to the differences in the data and in the analysis methodology. While \citet{smolecsniegowska16} conducted their study on OGLE-III light curves, we used the joined OGLE-II-III-IV data. In addition to the different pre-whitening procedures of the studies, the identification of the maximum in cases where the periodogram has two comparable peaks is very sensitive to the method applied. However, if the nonradial secondary modes are variable, and can significantly weaken or disappear over time, this can also partly contribute why we see such differences in our results.

\subsubsection{The 1.25-mode}
\label{subsubsec:1.2mode}

\begin{figure}
\begin{center}
	\includegraphics[width=0.85\columnwidth]{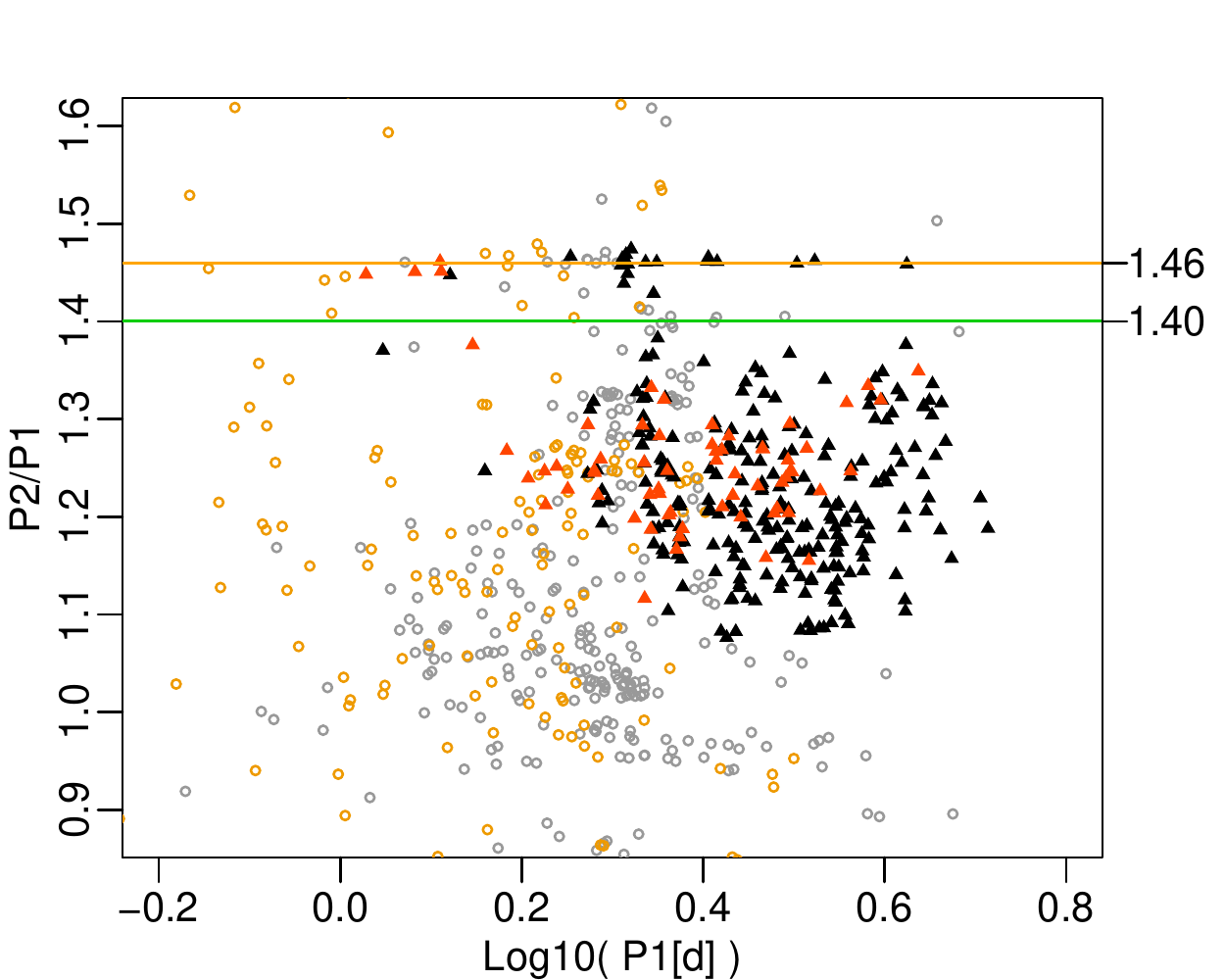}
\caption{Enlarged Petersen diagram in the region of the 1.25- and 1.46-mode. The colour code is the same as in the rightmost panel of Figure \ref{fig:petersen_full}. The plotted values of $P_2 / P_1$ are de-aliased (see text) for the unambiguously identified modes (i.e., filled triangles), while they are the ratio of the period corresponding to the highest secondary periodogram peak and the primary period for the non-identified modes (empty orange and grey circles). The orange line is at 1.465, indicating the approximate location of the 1.46-mode. The green line is at $1/0.714 \approx 1.4$, that is, the approximate location of a fundamental radial-mode frequency if it appeared as the secondary mode.}
\label{fig:petersen_12-mode}
\end{center}
\end{figure}


\begin{figure}
	\includegraphics[width=0.95\columnwidth]{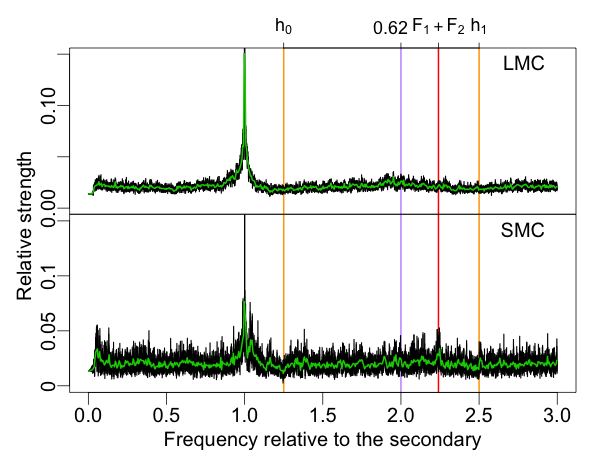}
\caption{Standardised average secondary periodograms, the average taken over those FO stars that have a secondary unambiguous non-aliased 1.25-mode in the LMC (top panel), and in the SMC (bottom panel). The vertical violet line denotes the position of the 0.62-mode. Other notation is the same as in Figure \ref{fig:oglenonrad_avgSecPgram}.}
\label{fig:funnyA_avgSecPgram}
\end{figure}

We have identified another sequence of modes where the ratio of the secondary and the primary period, $P_2/P_1$ is close to 1.25 ($F_2 / F_1 \approx 0.8$). The stars belonging to this family do not form such a sharp, well-defined almost-linear sequence as the 0.62-mode (cf. the lower panels of Fig.~\ref{fig:f1_f2_fo}). In the SMC, the $F_2 \approx 0.8 F_1$ sequence is relatively tight, although it appears to be more scattered than the 0.62-mode.  In the LMC, this narrow stripe is replaced by a wide, apparently curved, very densely populated band, which is predominant for primary frequencies $F_1 \in [0.2, 0.6] \, c/d$, but extends also to frequencies below and above this range. It is separated from the 1.46-mode  ($F_2 \approx 0.685 F_1$) below by a narrow near-empty region, and from the dense cluster of near-primary modes above by a wider gap in the interval $F_1 \in [0.45, 0.6] c/d$ in the LMC. 

The period ratio of this sequence is near to that of the first and second overtone, if the primary mode is in fact the second overtone, and the secondary mode is the first overtone. However, this possibility can easily be ruled out. Plotting the relative phases $\Phi_{21}$ and amplitudes $R_{21}$ of the dominant pulsation against $\log P_1$ 
and checking the period-luminosity relationship, 
the positions of these stars confirm unambiguously that their primary period is indeed a radial first overtone pulsation.

The period of this mode is very close to the double of that of the 0.62-mode (in other words, to the 1/2 subharmonic of its frequency). Thus, this mode might be the low-order non-radial mode at the origin of the 0.62-mode, directly observed at its fundamental frequency \citep{dziembowski16}. However, this is at odds with the explanation offered by  \citet{dziembowski16}: the contributions of the fundamental of this $l = 7,8,9$ mode from various surface elements of the star do not cancel perfectly in these objects. Moreover, the presence of the first harmonic (which corresponds to the 0.62-mode frequency) is uncertain and weak. The averaged standardised secondary spectra of unambiguous, non-aliased 1.25-mode stars in the LMC show on average increased power around the first harmonic ($\nu = 2$), but this increase is small, and the peak takes a flat, wide shape with the maximum at $\nu = 1.9$. There is no prominent peak at $\nu = 2$ in the average secondary periodograms in the SMC. Adding the spectra of aliased unambiguous detections to the mean periodogram, a more prominent peak emerges here for both Clouds. 

In both Clouds, the main peak of the average secondary periodograms of 1.25-mode stars (around $\nu = 1$) is much broader than the same main peak in 0.62- or 1.46-mode stars (compare with Figures  \ref{fig:oglenonrad_avgSecPgram} and \ref{fig:funnyB_avgSecPgram}). \citet{dziembowski16} suggested that this fundamental frequency may be more variable than its harmonic  the 0.62-mode (due to complex interactions within the multiplet), and the periodogram around it to be more complex. The observed width of the band in the $F_1-F_2$ plane where the 1.25-mode occurs may be at least partly due to this effect, either by identifying various components of the fundamental multiplet as main frequency, or by the broadening and distortion of the peak (cf. Figures \ref{fig:f1_f2_fo} and \ref{fig:petersen_12-mode}). The width of the 1.25-mode stripe can be explained also if the 1.25-mode is the product of more (or other) non-radial modes than just the $l=7,8,9$ ones. These assumptions can also contribute to explain why the region of increased power in the averaged secondary periodograms around $\nu = 2$ is so broad, and why the main peak at $\nu = 1$ is wider for the 1.25-mode than for the 0.62- or the 1.46-mode. 

As mentioned in Section \ref{subsubsec:0.6mode}, 14 of the SMC 1.25-mode objects were identified as (weak) 0.62-mode stars by \citet{soszynskietal15b}. In the spectrum of these, we find a usually weak peak at the 0.62-mode frequency. This raises the interesting possibility of a transition from one dominant harmonic to the other over time, if the 0.62- and the 1.25-mode is indeed in such a close relationship. A local kernel analysis using a two-mode model will give more insight into whether such transition happens indeed.

\subsubsection{The 1.46-mode}
\label{subsubsec:1.4mode}

\begin{figure}
	\includegraphics[width=0.95\columnwidth]{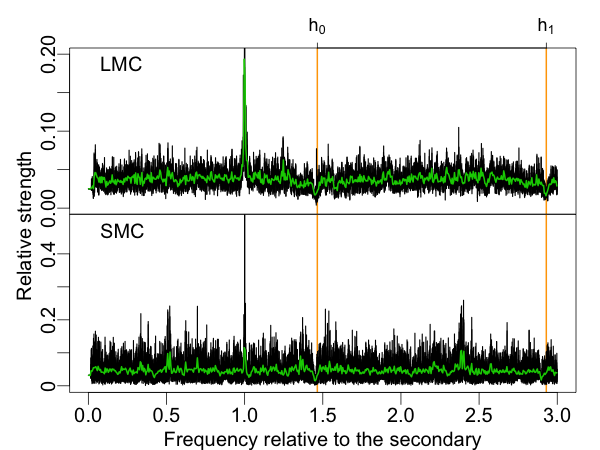}
\caption{Standardised average secondary periodograms, the average taken over those FO stars that have a secondary unambiguous 1.46-mode in the LMC (top panel), and in the SMC (bottom panel). Colour code is the same as in Figure \ref{fig:oglenonrad_avgSecPgram}.}
\label{fig:funnyB_avgSecPgram}
\end{figure}

%
We have identified another mode on the $F_1-F_2$ plane hitherto unknown in Cepheids, the sequence $F_2 \approx 0.69 F_1$. The line shape is much sharper, better defined than the loose, dispersed cluster of the 1.25-mode stars in the Petersen diagram, Figure  \ref{fig:petersen_12-mode}. The new mode is separated by a visible gap from the 1.25-mode (cf. Figures \ref{fig:f1_f2_fo}, \ref{fig:petersen_full} and \ref{fig:petersen_12-mode}). For comparison, Figure \ref{fig:petersen_12-mode} shows where a weak radial fundamental mode would approximately fall if we detected it as a secondary.

The assumption that this is another, independent pulsation mode is supported by the fact that \citet{netzeletal15} found a weak mode in first overtone RR Lyrae pulsators (RRc variables) with a very similar period ratio. The origin of this oscillation in RRc stars is unknown, because according to theory, the modes that could explain such long periods are stable in the relevant conditions. An alternative explanation was proposed by \citet{pietrzynskietal12}, advancing that these objects may be giants stripped of their outer layers through binary evolution. The existence of a very similar mode in first overtone Cepheids, further increasing the already remarkable similarity between RRc and FO Cepheid stars, suggests a common origin in pulsation theory for the two types of classical pulsators.

For the averaged periodograms in Figure \ref{fig:funnyB_avgSecPgram}, we used all unambiguous detections, aliased or not. Even so, the average in the SMC is based on only four stars (in the LMC, we have 19 stars). It shows little structure, apart from the dips due to pre-whitening in the LMC, and a broad peak between 2.25 and 2.4 in the SMC. However, this average periodogram is based on only four objects, so the presence of this peak may be due simply to small-sample random effects.

\section{Characterisation of the secondary mode populations}
\label{sec:char}

\subsection{Primary periods}

\begin{figure}
\begin{center}
	\includegraphics[width=0.95\columnwidth]{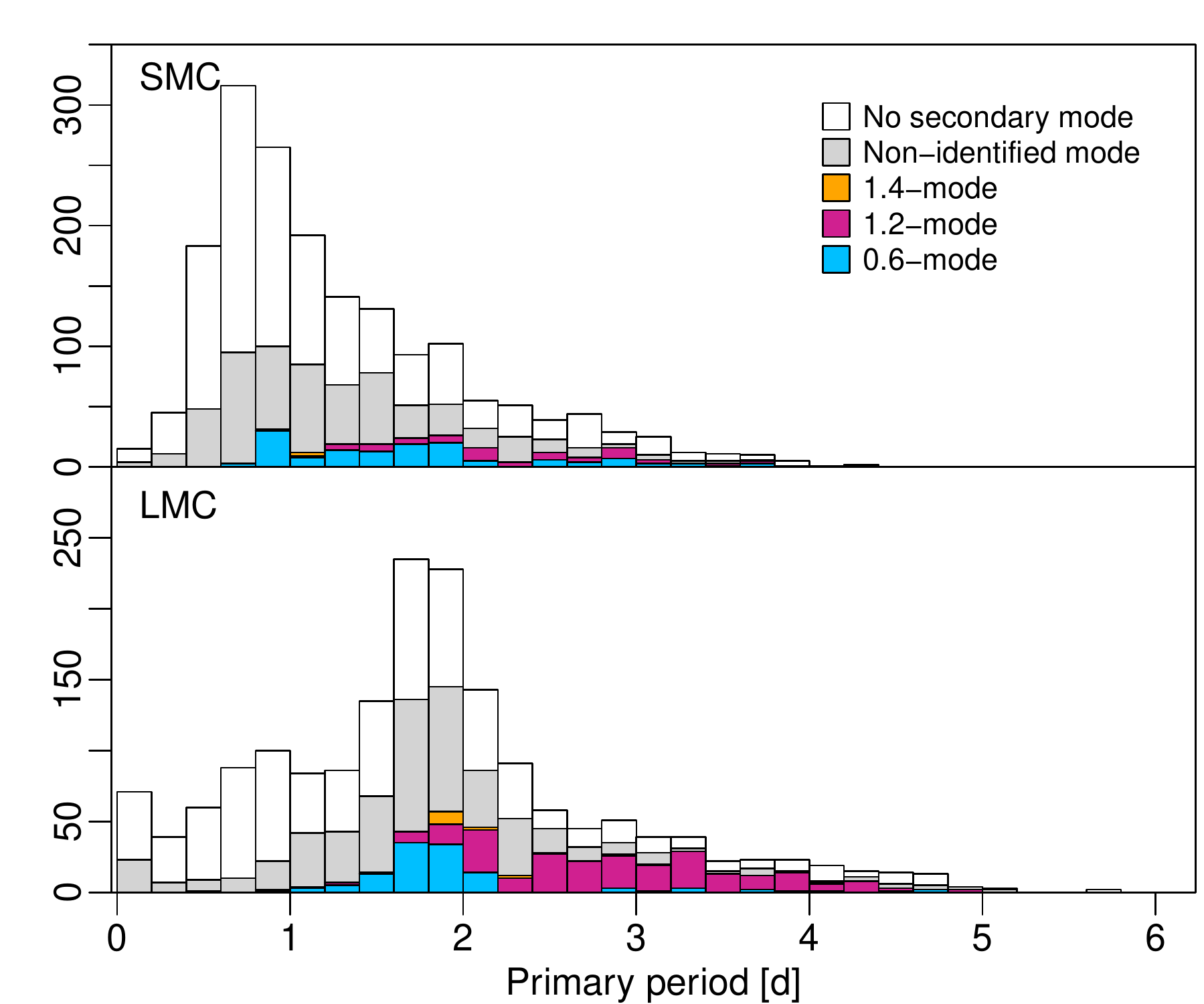}
\caption{Histogram of the primary periods of the FO Cepheids in the sample. The bars are subdivided according to the secondary mode content: the number of unambiguous 0.62-modes is indicated in blue, that of unambiguous 1.25-modes in purple, that of unambiguous 1.46-modes in orange, that of other, unidentified significant modes in grey. Top panel: SMC, bottom panel: LMC.}
\label{fig:hist_primperiods}
\end{center}
\end{figure}

Figure \ref{fig:hist_primperiods} shows how the occurrence rate of the secondary modes depends on the primary periods. In the SMC, the 0.62-mode occurs in a broader period range than in the LMC, and while in the SMC it can be found in stars with period as short as $0.7\, d$ or as long as $3.5\, d$, in the LMC it does not occur below $P_1 = 1\, d$ or above $2.2\, d$. The 1.25-mode stars behave differently: they occupy a broader primary period range in the LMC than in the SMC, populating the period range between $2.2\, d$ and $5\, d$ in the LMC where there are no 0.62-mode stars. In the SMC, both modes occur together with the longest primary periods, although the 0.62-mode appears slightly more frequent. Qualitatively, the 0.62-modes associated with relatively longer periods (with respect to the distribution in the concerned Cloud) are missing in the LMC, and apparently, their place is taken over by the 1.25-mode. This further corroborates the putative existence of a close link between these two modes. 

\subsection{Secondary amplitudes} \label{subsec:secamp}

\begin{figure}
\begin{center}
	\includegraphics[width=\columnwidth]{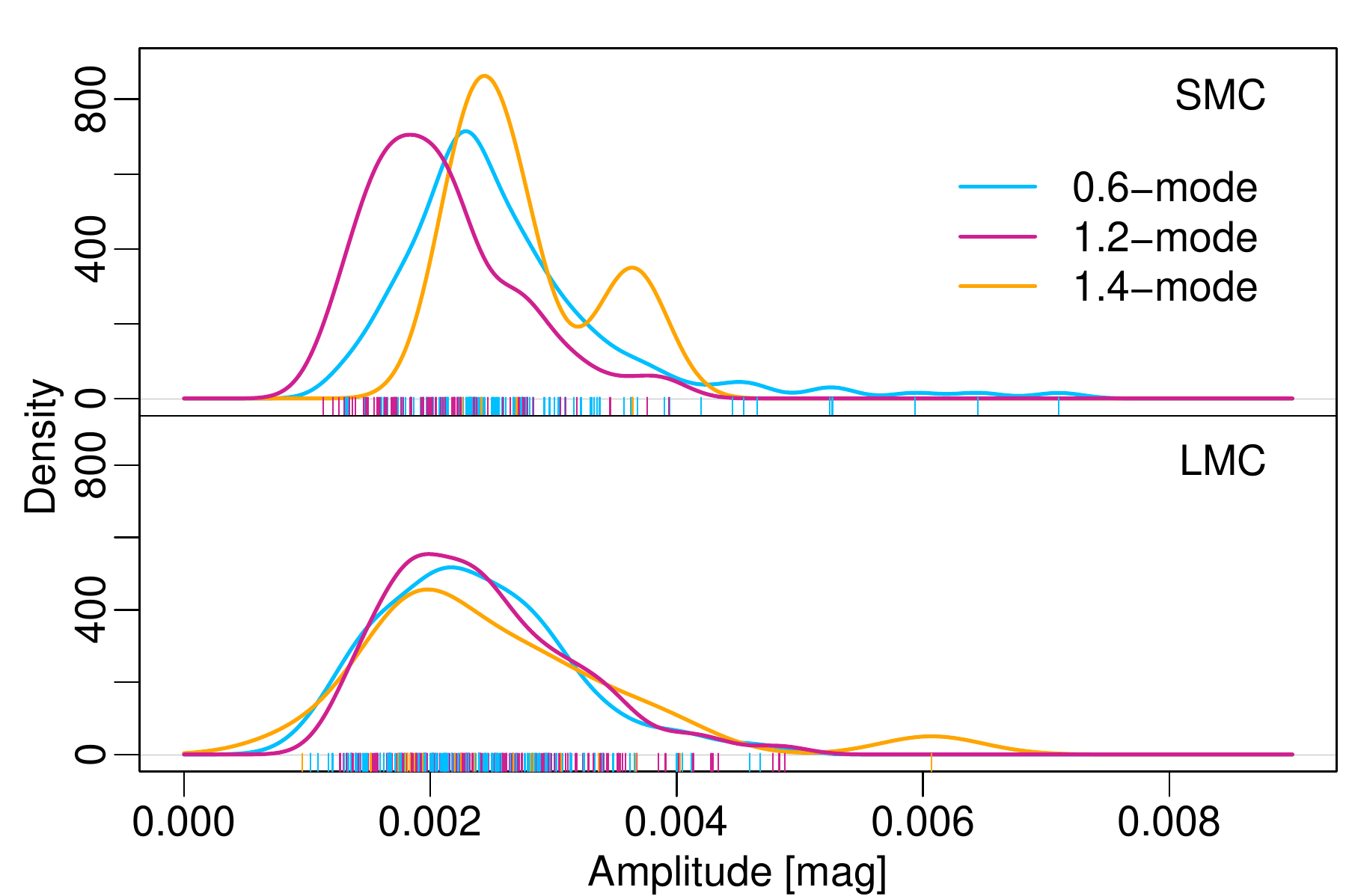}
\caption{Kernel density estimate of the secondary amplitude distribution of OGLE FO Cepheids with (using only unambiguous) 0.62-mode (blue), 1.25-mode (purple) and 1.46-mode (orange). The rugplot at the bottom shows the data, secondary mode types distinguished by colour. Top panel: SMC, bottom panel: LMC.}
\label{fig:hist_secamps}
\end{center}
\end{figure}

The distribution of the amplitudes of the three identified secondary modes is presented in Figure \ref{fig:hist_secamps} (a single 0.62-mode star with an exceptionally high amplitude of 15 mmag in the LMC is omitted from the plots). The distributions of the 0.62- and 1.25-mode in the LMC are similar, somewhat skewed bell-shapes, located at close-by average amplitude values. In the SMC, they are somewhat different: the 1.25-mode has an amplitude distribution concentrated at lower values, while the 0.62-mode has on average higher amplitudes in the SMC than the 1.25-mode. The amplitude distribution of the 1.46-mode is difficult to judge, since it is based on 19 stars in the LMC and only 4 in the SMC, though it appears to extend to relatively high amplitudes (4 mmag and above) with a somewhat higher probability than the 0.62- and the 1.25-mode. The lower limit of the distributions is set by the detection limit, somewhat above 1 mmag.  

Following \citet{smolecsniegowska16}, we also checked plots of the primary amplitude versus the period ratio $P_2/P_1$ and the amplitude ratios $A_2/A_1$ for all three secondary mode types in both Clouds. Apart from a weak negative correlation between the period ratio and the primary amplitude for SMC 0.62-mode stars noted also by them, we did not find any notable relationships between any of these parameters for the other modes, either in the SMC or in the LMC. We do not show these figures.

\subsection{Period-luminosity relationship and colour-magnitude diagram}

\begin{figure}
\begin{center}
	\includegraphics[width=\columnwidth]{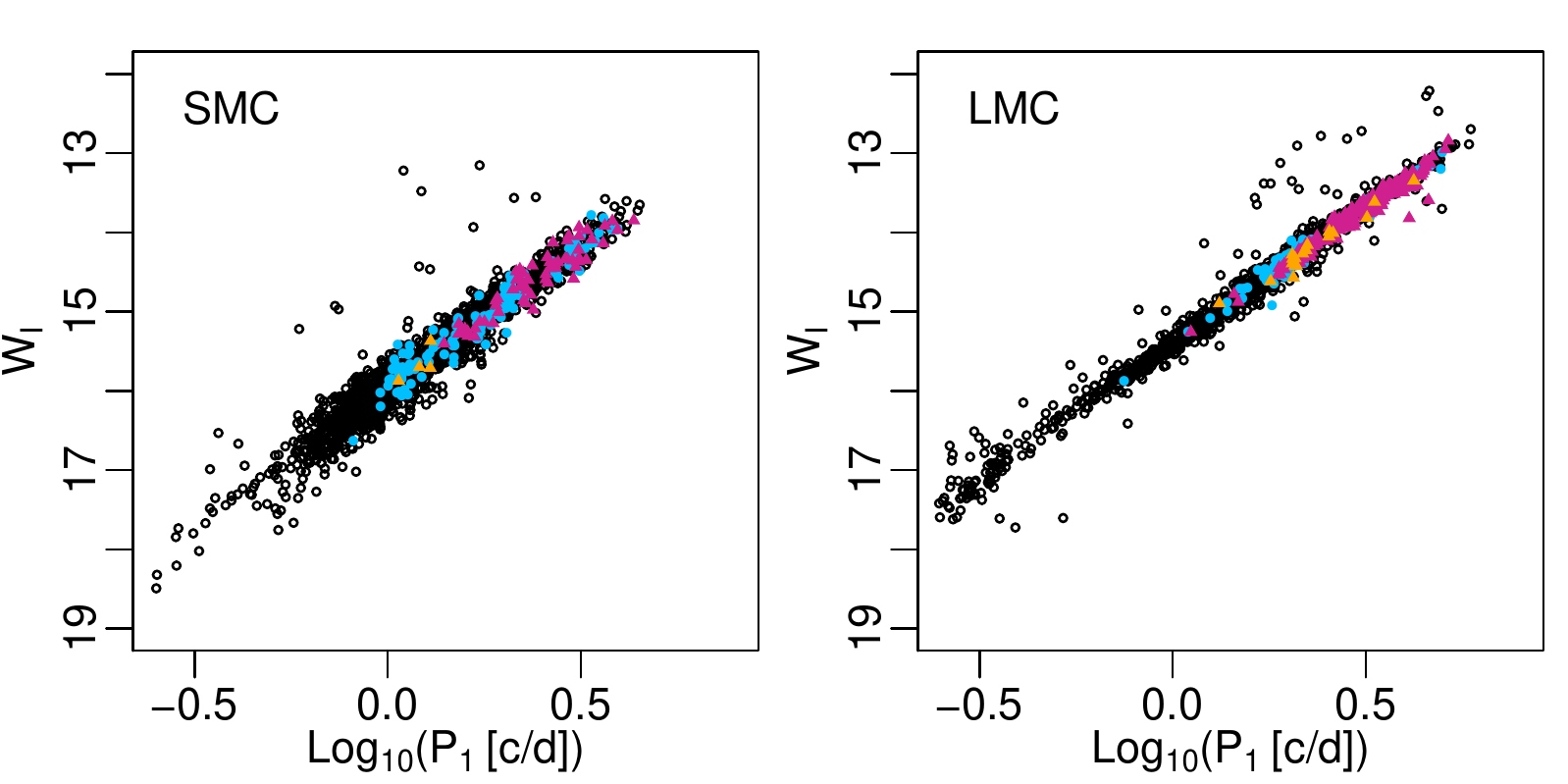}
\caption{Period-luminosity relationship of classical first-overtone Cepheids in the SMC (left panel) and in the LMC (right panel). The three secondary mode types are indicated by blue dots (0.62-mode), purple triangles (1.25-mode) and orange triangles (1.46-mode).}
\label{fig:prl}
\end{center}
\end{figure}

\begin{figure}
\begin{center}
	\includegraphics[width=\columnwidth]{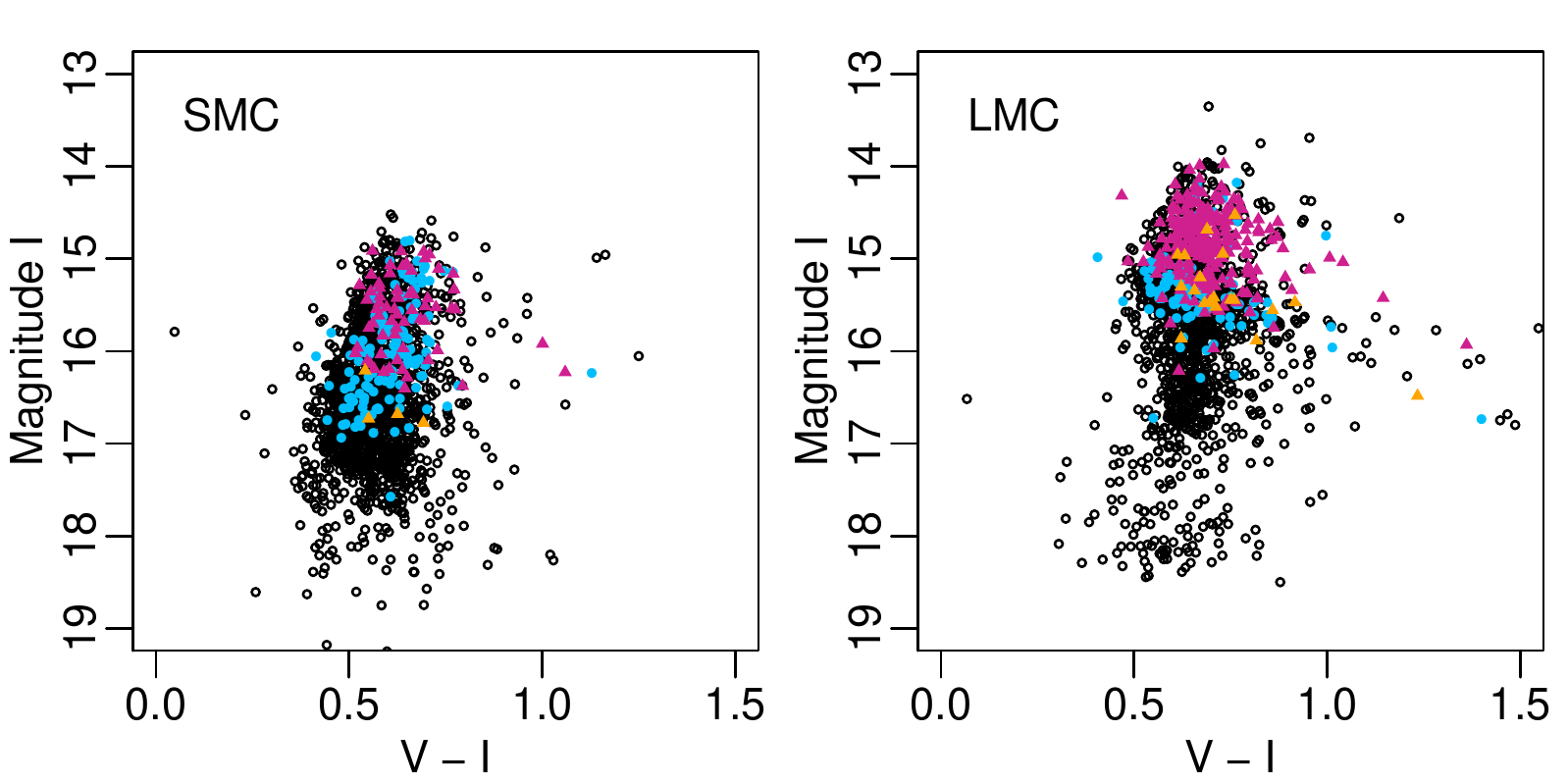}
\caption{Colour-magnitude diagram of classical first-overtone Cepheids in the SMC (left panel) and in the LMC (right panel). The three secondary mode types are indicated by blue dots (0.62-mode), purple triangles (1.25-mode) and orange triangles (1.46-mode).}
\label{fig:cmd}
\end{center}
\end{figure}

Figure \ref{fig:prl} shows the reddening-free Wesenheit index $W_I = I - 1.55(V-I)$ versus the logarithm of the primary period in the sample. Cepheids with different secondary modes occupy different period ranges, as Figure \ref{fig:hist_primperiods} demostrated. Within their own period range however, they follow the general luminosity distribution of FO Cepheids.  Figure \ref{fig:cmd} yields a similar impression about the colour-magnitude diagram. Since the different modes occur with different typical primary periods, the distribution of their mean magnitude $I$ also differs, and thus, the three modes occupy different, yet overlapping apparent magnitude ranges. Their $V-I$ colours, on the contrary, follow the general distribution of first overtone Cepheids: no particular structure (or favoured location within the instability strip) is seen in the colour-magnitude diagram.

\section{Conclusions}
\label{sec:concl}

We have created an inventory of weak secondary modes among classical Cepheids labelled as single-mode in the OGLE-II-III-IV data of the Magellanic Clouds. Recently, many of these were found to have modulated primary pulsation \citep[][Paper I]{suvegesanderson17}, so to achieve a more complete list than is available so far, we used the local kernel regression method (\citetalias{suvegesanderson17}) for pre-whitening after modifying it for adaptive kernel widths.  This method models better an irregularly modulated primary, and thus reduces the unexplained, noise-like variations beside a weak secondary periodic signal in the residual time series, although it also removes potential signals of interest if their period is very close to the primary period. 

The secondary period search, performed on the residuals using the generalised least squares method \citep{zechmeisterkurster09}, yields mostly low-amplitude (in large majority below 5 mmag) secondary modes. These are nearly uniformly distributed over the $F_1 - F_2$ plane for the fundamental Cepheids, and exhibit an intriguing rich structure for first overtone Cepheids. Simulations (cf. Section \ref{subsec:detectability}) suggest that the lower limit of the amplitudes found, somewhat above 1 mmag, is imposed by the OGLE photometric precision error, the time sampling and the limitations of the kernel pre-whitening. It seems likely that this $\sim\!\! 1$ mmag limit is not an intrinsic physical limit of the excited modes, and there might exist many more weak oscillations in classical Cepheids both in the identified modes and in ones unknown so far because of their weakness.  The secondary mode content of fundamental Cepheids, with their patternless Petersen diagram and $F_1 - F_2$ plane, appears different from the fundamental mode RR Lyrae stars, which have a wealth of secondary modes clustering in several characteristic groups \citep{molnaretal17}. The pattern of secondary modes in first overtone Cepheids shows some features similar to that of RRc stars, such as the presence of the 0.62-mode or the (newly discovered) 1.46-mode, but there are also differences, such as the substructure of the Cepheid 0.62-mode or the existence of the 1.25-mode, neither of which has yet been detected in RRc stars.  

Counting only unambiguously identified modes, we have detected 139 (120) 0.62-mode objects in the SMC (LMC). Including all ambiguous 0.62-modes, these numbers increase to 271 in the SMC and 209 in the LMC. The analysis of these stars confirmed previous knowledge \citep{smolecsniegowska16}. Moreover, the adaptive kernel pre-whitening led to the discovery of at least two new modes among the first overtone Cepheids: a mode at approximately the 1/2 subharmonic of the 0.62-mode (1.25-mode), and another one at $P_2 = 1.46 P_1$ (1.46-mode). 

Both of these modes are extremely interesting for the  modelling of the pulsation and the stellar structure of Cepheids. If we accept the interpretation of the 0.62-mode as the first harmonic of the $l \in \{7,8,9\}$ non-radial modes \citep{dziembowski16}, a potential explanation for the 1.25-mode is that it is the individually existing fundamental of the same non-radial modes. Different conditions in the two Clouds might favour the dominant appearance of one or the other. The complementary distribution of the two modes in the Clouds lends support to this possibility: the 0.62-mode has been known to be more frequent in the SMC than in the LMC \citep{soszynskietal15b,smolecsniegowska16}, while the 1.25-mode is far more frequent in the LMC than in the SMC. The 0.62-mode is clearly dominant in both Clouds with primary periods shorter than about 2 days, but its dominance disappears above this limit. With primary periods longer than 2.5 days, the LMC has an overwhelmingly prevalent population of  1.25-modes (more than half of all FO Cepheids show it in this primary period range). This 1.25-mode population associated with long primary periods is almost absent in the SMC. The location of the 1.25-mode in the  $F_1 - F_2$ plane and in the Petersen diagram is broad, in particular in the LMC, which might be related to the predicted complexity of these modes, and potentially, to the presence of other orders $l$. 


The very rare 1.46-mode is another new discovery in Cepheids. With its period ratio of 1.465, it appears to parallel a mode with period ratio 1.458 in first overtone RR Lyrae pulsators \citep{netzeletal15}. This mode, similarly to the 1.25-mode, is more frequent in the LMC than in the SMC, which might in part be due to the lower detection efficiency in the SMC (cf. \ref{subsec:detectability}). In the SMC, the 1.46-mode occurs together with  shorter primary periods than in the LMC. 

In addition to these modes, there are many stars with primary periods mostly in the range of $[1.7, 2] $ days that show near-primary secondary frequencies with period ratios between 0.95 and 1.1. A subset of these, with a period ratio of about 0.95, appears in the $F_1 - F_2$ plot to form a near-linear sequence. However, since the local kernel pre-whitening may influence the detection probability and bias the fit of these modes, we do not analyse this group in this paper. 

At present, though for the 0.62-mode theory has proposed a plausible interpretation, there is no satisfactory  explanation for the 1.46-mode or the 1.25-mode. The inventory presented in this study raises many theoretical questions. What is the origin of the large group termed here 1.25-mode, and why does this group occupy such a broad and complex region in the $F_1 - F_2$ plane? Are the 1.25- and the 0.62-mode indeed related, and if so, in what way? How are the marked differences in the mode populations in the LMC and the SMC related to overall differences of the Clouds (e.g., metallicity)? What is the origin of the 1.46-mode? How can the similarities between first overtone RR Lyrae and Cepheid pulsators help understanding the physical origins of these modes in the stars,  why do the fundamental modes of the two classical pulsators appear different, and why is the 1.25-mode absent (or as yet unobserved) in RR Lyrae stars? Further research in this direction, powered e.g. by the soon-to-be-launched TESS mission, will likely provide answers to such questions, thereby furthering our understanding of stellar pulsations and structure.

\section*{Acknowledgements}
We thank the anonymous referee for constructive questions and suggestions, in particular regarding differences and similarities of additional modes found in RR Lyrae and classical Cepheid stars depending on their primary pulsation mode (fundamental, first overtone).




\bibliographystyle{mnras}
\bibliography{bibfileAstro,bibfileOrig} 









\bsp	
\label{lastpage}
\end{document}